\documentclass[12pt, draftclsnofoot, onecolumn]{IEEEtran}

\title{Effect of LOS/NLOS Propagation on Ultra-Dense Networks}
\author{{\normalsize Carlo Galiotto$^1$,~~Nuno K. Pratas$^2$,~~Linda Doyle$^1$,~~Nicola Marchetti$^1$
\\
	1) CTVR, Trinity College, Dublin, Ireland \\
	2) Department of Electronic Systems,
	Aalborg University (AAU), Denmark \\ 1) \{galiotc,~Linda.Doyle,~marchetn\}@tcd.ie,~~~~2) nup@es.aau.dk.}

}

\usepackage[dvips]{graphicx}
\usepackage[T1]{fontenc}
\usepackage{amsmath}
\usepackage{amssymb}
\usepackage{amsthm}
\usepackage{amssymb}
\usepackage{amsfonts}
\usepackage{cite}
\usepackage[nolist]{acronym}
\usepackage{microtype}
\usepackage{tabularx}
\usepackage{multirow}
\usepackage{latexsym}
\usepackage{eurosym}
\usepackage{here}
\usepackage{soul} 		
\usepackage{epsfig}
\usepackage{epstopdf}
\usepackage{color}      

\usepackage{algorithm}
\usepackage[noend]{algpseudocode}

\usepackage[caption=false,font=footnotesize]{subfig}
\usepackage{float}                    
\usepackage{lettrine}                

\newcommand{\figureSize}{0.55}

\newtheorem{definition}{Definition}

\newtheorem{result}{Result}

\begin{document}
	
\bstctlcite{IEEEexample:BSTcontrol}
\maketitle
\begin{abstract}

This paper aims at investigating the achievable performance and the issues that arise in ultra-dense networks (UDNs), when the signal propagation includes both the Line-of-Sight (LOS) and Non-Line-Of-Sight (NLOS) components. Backed by an analytical stochastic geometry-based model, we study the coverage, the Area Spectral Efficiency (ASE) and the energy efficiency of UDNs with LOS/NLOS propagation. We show that when LOS/NLOS propagation components are accounted for, the network suffers from low coverage and the ASE gain is lower than linear at high base station densities. However, this performance drop can partially be attenuated by means of frequency reuse, which is shown to improve the ASE vs coverage trade-off of cell densification, provided that we have a degree of freedom on the density of cells. In addition, from an energy efficiency standpoint, cell densification is shown to be inefficient when both LOS and NLOS components are taken into account. Overall, based on the findings of our work that assumes a more advanced system model compared to the current state-of-the-art, we claim that highly crowded environments of users represent the worst case scenario for ultra-dense networks. Namely, these are likely to face serious issues in terms of limited coverage.
		
\end{abstract}




\section{Introduction}
\label{sect:intro}

There is a common and widely shared vision that next generation wireless networks will witness the proliferation of small-cells. As a matter of fact, researchers foresee network densification as one of key enablers of the 5\textsuperscript{-th} generation (5G) wireless networks \cite{Hwang2013,Bhushan2014}. Although it refers to a concept rather than being a precise definition, the term \textit{ultra-dense networks} is used to describe networks characterized by a massive and dense deployment of small-cells, in which the amount of base stations may grow up to a point where it will exceed the amount of user devices \cite{Park2014}.

As wireless networks evolve, performance requirements for the new technology are becoming more and more stringent. In fact, the 5G requirements are set to a data rate increase up to a 1000-fold with respect to current 4G systems\cite{Bhushan2014}, as well as for high energy efficiency \cite{Ericsson2015} in order to limit the energy expenditure of network operators. Supported by recent results \cite{Bhushan2014,Andrews2011}, cell densification as been put forward as the main enabler to achieve these target data rates. For example, the authors in~\cite{Andrews2011} have shown that the throughput gain is expected to grow linearly with the density of base stations per area; this is a result of the simplified system model used during the investigation. Namely, the assumption of a single slope path-loss model and that all base stations in the network are active.

Nonetheless, further work on cell densification has shown that, under less ideal assumptions, network performance may be different than what predicted in~\cite{Andrews2011}. In particular, when different path-loss models than single slope are used, the actual performance of cell densification is less optimistic than what estimated with single slope path-loss \cite{Zhang2014,Bai2014,Galiotto2015,Ding2015}. In addition, if the base station density increases beyond the user density, like in a typical ultra-dense scenario, the network has been shown to experience a coverage improvement at the expense of a limited throughput gain \cite{ChangLi2014,Park2014}. This implies that a larger number of BSs will need to be deployed to meet a given data rate target, translating on higher network infrastructures costs.

\subsection{Related Work} 
\label{sub:related_work}

In recent years, stochastic geometry has been gradually accepted as a mathematical tool for performance assessment of wireless networks. In fact, one of the most important contribution to the study of cell densification can be found in~\cite{Andrews2011}, where the authors proposed a stochastic geometry-based framework to model single-tier cellular wireless networks; by assuming a single slope path loss model, the authors have observed the independence of the Signal-to-Interference-plus-Noise-Ratio (SINR) and Spectral Efficiency (SE) from the BS deployment density, with the main consequence being the linear dependence of the ASE on the cell density.

Nevertheless, when the assumption of single-slope path-loss is dropped, it emerges that SINR, ASE and coverage exhibit a different behaviour than what was found in~\cite{Andrews2011}. The authors in~\cite{Zhang2014,Bai2014a} considered propagation models for millimeter waves. In~\cite{Zhang2014}, the authors extended the stochastic geometry framework proposed in~\cite{Andrews2011} to a multi-slope path loss model. The authors in~\cite{Bai2014a} developed a stochastic geometry framework for path-loss including Line-of-Sight (LOS) and Non-Line-of-Sight (NLOS) propagation for millimeter-waves. The effect of NLOS propagation on the outage probability has been studied in~\cite{Bai2014}, where the authors propose a function that gives the probability to have LOS at a given point depending on the distance from the source, on the average size of the buildings and on the density of the buildings per area. A stochastic geometry-based framework to study the performance of the network with a combined LOS/NLOS propagation for micro-waves can be found in some previous work of ours \cite{Galiotto2015} and in \cite{Ding2015}.
The common picture that emerges from the work  \cite{Zhang2014,Bai2014,Galiotto2015,Ding2015} is that of a non-linear behavior of the ASE with the cell density; overall, as a consequence of a different propagation model than the single slope path-loss, the spectral efficiency and coverage of the network do actually depend on the base station density.
However, all these studies are based on the assumption of fully loaded networks, i.e., all the base stations are active and have at least one user to serve; thus, the applicability of the work above is limited to networks in which the density of base stations is lower than that of the users. In our paper, we broaden the study of network densification to the the case where there is no such a constraint in terms of BS density, i.e.,  we also tackle partially loaded networks, in which some base stations might be inactive.

Work on stochastic geometry for partially loaded networks has been advanced in \cite{Park2014,SeunghyunLee2012,Dhillon2013,ChangLi2014}.  The authors in \cite{SeunghyunLee2012} studied the coverage in single tier networks, while multi-tier networks are addressed in \cite{Dhillon2013}. An analysis of the area spectral efficiency of partially loaded networks has been carried out in \cite{Park2014}, while in \cite{ChangLi2014} the authors  have extended the stochastic geometry-based model further to include multi-antenna transmission, and have also assessed the energy efficiency. Overall, the authors in \cite{SeunghyunLee2012,Dhillon2013,ChangLi2014} have shown that the network coverage improves as the base station density increases beyond the user density; this is paid in terms of a lower throughput gain, which grows as a logarithmic function of the cell density. Nonetheless, the authors in \cite{SeunghyunLee2012,Dhillon2013,ChangLi2014} modeled the propagation according to a single slope path-loss model and did not investigate the effect of LOS/NLOS propagation in partially loaded networks.

To the best of our knowledge, currently there is no work that addresses ultra-dense scenarios with path-loss models different than the single slope for both fully and partially loaded networks. As a result of the combined effect of the path-loss model and of the partial load in ultra-dense networks, until now the behavior of the Area Spectral Efficiency (ASE), coverage, and energy efficiency as the networks become denser, was unknown.


\subsection{Our Contribution} 
\label{sub:our_contribution}

This paper seeks to investigate the cell densification process in ultra-dense networks and evaluate the effect of LOS/NLOS propagation  on performance metrics such as coverage, spectral efficiency, area spectral efficiency, and energy efficiency. Specifically, we use \emph{ultra-dense networks} as a term to refer to those networks characterized by a very high density of base stations and that include both the cases of \emph{fully loaded networks} (i.e., networks in which all the base stations are active) and  \emph{partially loaded networks} (i.e., networks in which some base stations might be inactive and not transmit to any user). Overall, the major contributions of our work can be summarized in the following points:

\textbf{1) Stochastic geometry-based model for ultra-dense networks with LOS/ NLOS propagation:} We propose a model based on stochastic geometry that allows us to study the Signal-to-Interference-plus-Noise-Ratio (SINR) distribution, the spectral efficiency and the area spectral efficiency of ultra-dense networks where the propagation has LOS and NLOS components. We build on previous work \cite{Andrews2011} and we adapt the model proposed by Andrews et al. to the case of LOS/NLOS propagation. In addition, our framework takes also into account the partial load of ultra-dense networks, in which a fraction of base stations may be inactive.

\textbf{2) Study of cell densification, partial load and frequency reuse in networks where signal follows LOS/NLOS propagation:} First, we investigate the effect of network densification on performance metrics such as SINR, spectral efficiency and area spectral efficiency in networks where the path-loss follows the LOS/NLOS propagation. In particular, the ASE gain becomes lower than linear at high cell densities, meaning that a larger number of BSs would be necessary to achieve a given target with respect to the case of single slope path loss. Moreover, the network coverage drops drastically as the BS density increases. Then, we show that the performance drop due to LOS/NLOS propagation gets mitigated by the usage of frequency reuse or if the base station density exceeds the user density, as it is likely to occur in ultra-dense networks. To the best of our knowledge, the combined effect of LOS/NLOS propagation and partial load/frequency reuse has not been addressed before.

\textbf{3) Investigation on the minimum transmit power per BS and energy efficiency for networks where signal propagates according to LOS/NLOS  path-loss:} As the cell density increases, the transmit (TX) power per base station can be lowered. We evaluate the minimum TX power per BS such that the network is guaranteed to be in the interference-limited regime, in which case the performance is not limited by the TX power. Second, we make use of the TX power to determine the energy efficiency of the network when the propagation has LOS and NLOS components. We show that the energy efficiency with LOS/NLOS propagation drops considerably with respect to the case of single slope path-loss, making cell-densification costly for the network from an energetic stand-point. We further extend the study of the energy efficiency to frequency reuse and partial load.
\vspace{-3mm}

\subsection{Paper Structure} 
\label{sub:paper_structure}

The remainder of this paper is organized as follows. In Section~\ref{sect:SystemModel} we describe the system model. We show our formulation for computing the SINR, SE and ASE in Section~\ref{sect:SINR_SE_ASE} and we address the energy efficiency in Section \ref{sec:energy_Efficiency}. In Section~\ref{sect:results} we present and discuss the results while the conclusions are drawn in Section \ref{sect:conclusions}.
\vspace{-3mm}


\section{System model}\label{sect:SystemModel}

In this paper we consider a network of small-cell base stations deployed
according to a homogeneous and isotropic Spatial Poisson Point Process (SPPP), denoted as $\Phi\subset \mathbb{R}^2$, with intensity $\lambda$. Further, we assume that each Base Station (BS) transmits with an isotropic antenna and with the same power, $P_{\mathrm{TX}}$, of which the value is not specified, in order to keep our model general and valid for different base station classes (e.g., micro-BSs, pico-BSs, femto-BSs); we focus our analysis on the downlink.
\vspace{-3mm}

\subsection{Channel model} \label{subsect:ChannelModel}

In our analysis, we considered the following path loss model:
\begin{equation}
	\mathrm{PL}(d)=\begin{cases}
		K_{\mathrm{L}}d^{-\beta_{\mathrm{L}}} & \text{with probability}\; p_{
\mathrm{L}}(d),\\
		K_{\mathrm{NL}}d^{-\beta_{\mathrm{NL}}} & \text{with probability}\:1-p_{
\mathrm{L}}(d),
	\end{cases}\label{eq:propag_outdoor}
\end{equation}
where $\beta_{\mathrm{L}}$ and $\beta_{\mathrm{NL}}$ are the path-loss
exponents for LOS and NLOS propagation, respectively; $K_{\mathrm{L}}$ and $K_{\mathrm{NL}}$ are the signal attenuations at distance $d=1$ for LOS and NLOS propagation,\footnote{The parameters $K_{\mathrm{L}}$ and $K_{\mathrm{NL}}$
can either refer to the signal attenuations at distance $d=1$ m or $d=1$ km;
this depends on the actual values given for the parameters of the channel
model.} respectively; $p_{\mathrm{L}}(d)$ is the probability of having LOS as a function of the distance $d$. The model given in \eqref{eq:propag_outdoor}
is used by the 3GPP to model the LOS/NLOS propagation, for example, in
scenarios with Heterogeneous Networks~\cite[Table A.2.1.1.2-3]{3GPP36814}.
The incorporation of the NLOS component in the path loss model accounts for
possible obstructions of the signal due to large scale objects (e.g. buildings ), which will result in a higher attenuation of the NLOS propagation compared to the LOS path. We further assume that the propagation is affected by Rayleigh fading, which is exponentially distributed $\sim \exp(\mu)$.

Regarding the shadow fading, it has been shown that in networks with a
deterministic, either regular or irregular, base station distribution
affected by log-normal shadow fading, the statistic of the propagation
coefficients converges to that of a network with SPPP distribution as the
shadowing variance increases \cite{Blaszczyszyn2015}. In other words, this
SPPP intrinsically models the effect of shadow fading.
\vspace{-3mm}

\subsection{LOS probability function}\label{subsect:choosing_p_L}

To ensure that our formulation and the outcomes of our study are general and
not limited to a specific LOS probability pattern, we consider two different
LOS probability functions. The first one, which is proposed by the 3GPP~\cite[Table A.2.1.1.2-3]{3GPP36814} to assess the network performance in pico-cell scenarios, is given below:
\begin{equation}\label{eq:3GPP_p_L}
	p_{\mathrm{L,3G}}(d) = 0.5-\min\big(0.5,5e^{-\frac{d_0}{d}}\big)+\min\big(0.
5, 5e^{-\frac{d}{d_1}}\big),
\end{equation}
with $d_0$ and $d_1$ being two parameters that allow \eqref{eq:3GPP_p_L} to
match the measurement data. Unfortunately, this function is not practical for an analytical formulation. Therefore, we chose to approximate it with a more
tractable one, namely:
\vspace{-2mm}%
\begin{equation}\label{eq:Our_p_L}
	p_{\mathrm{L}}(d) = \exp\left( -(d/L)^2\right),
\end{equation}
where $L$ is a parameter that allows~\eqref{eq:Our_p_L} to be tuned to match~\eqref{eq:3GPP_p_L}. The second function is also suggested by the 3GPP~\cite[Table A.2.1.1.2-3]{3GPP36814} and is given below:
\begin{equation}\label{eq:p_L_exp}
	p_{\mathrm{L}}(d) = \exp(-d/L).
\end{equation}
From a physical stand point, the parameter $L$ can be interpreted as the LOS likelihood of a given propagation environment as a function of the distance.

\subsection{User distribution, fully and partially loaded networks }\label{subsect:user_distribution}

In our model, we always assume that: (i) the users are uniformly distributed; (ii) that each user's position is independent of the other users' position; and  (iii) each user connects only to one base station, the one that provides the strongest signal. We denote by $\lambda_{\mathrm{U}}$ the density of users per area; whenever we consider a finite area $A$, $N_{\mathrm{U}}$ indicates the average number of users in the network. We also assume the users are served  with full buffer, i.e., the base station has always data to transmit to the users and make full use of the available resources.

Depending on the ratio between the density of users and the density of base
stations, we distinguish between two cases, namely, fully loaded and
partially (or fractionally) loaded networks. By fully loaded networks we
refer to the case where each BS has at least one user to serve. With
reference to a real scenario, fully loaded networks model the case where
there are many more users than base stations, so that each base station
serves a non-empty set of users.

However, when the density of users is comparable with or even less that of
base stations, some base stations may not have any users to serve and would
then be inactive, meaning that they do not transmit and do not generate
interference. When this occurs, we say that the network is partially loaded.
The network can be modelled as partially loaded to study those scenarios
characterized by high density of base stations and, in particular, scenarios
where the density of base stations exceeds the density of users, such as in
ultra-dense networks.

To define formally the concepts of fully and partially loaded networks, we
first need to introduce another concept, which is the \textit{probability a
base station being active}.

\begin{definition}[Probability of a base station being active]\label{def:probability_of_activity}
The probability of a base station being active, denoted as $p_{\mathrm{A}}$,
is the probability that a base station has at least one user to serve. This
event implies that the base station is active and transmits to its users.
\end{definition}

\begin{definition}[Fully loaded and partially loaded networks]\label{def:fully_partially_loaded}
The network is said to be fully loaded if $p_A=1$; the network is said to be
partially or fractionally loaded if $p_A<1$.
\end{definition}


\section{SINR, spectral efficiency and ASE}\label{sect:SINR_SE_ASE}

In this section we propose an analytical model to compute the SINR Complementary Cumulative Distribution Function (CCDF), which allows us to asses key performance metrics such as coverage, spectral efficiency and ASE.

\subsection{Procedure to compute the SINR CCDF}

In order to compute the SINR tail distribution (i.e., the Complementary CDF), we extend the analytical framework first proposed in~\cite{Andrews2011} so that to include the LOS and NLOS components. From the Slivnyak's Theorem~\cite[Theorem8.10]{Haenggi2013}, we consider the~\emph{typical user} as the focus of our analysis, which for convenience is assumed to be located at the origin. The procedure is composed of two steps: (i) we compute the SINR CCDF for the typical user conditioned on the distance from the user to the serving base station, denoted as $r$; (ii) using the PDF of the distance from the closest BS $f_{r}(R)$, which corresponds to the serving BS, we can average the SINR CCDF over all possible values of distance $r$.

Let us denote the SINR by $\gamma$; formally, the CCDF of $\gamma$ is computed
as:
\begin{equation}\label{eq:SINR_general_def}
\mathrm{P}\left[\gamma>y\right]=\mathrm{E_{r}}\big[\mathrm{P} \left[\gamma>y|r\right]\big]
= \int _{0}^{+\infty} \mathrm{P}\left[\gamma>y|r=R\right]f_{r}(R)\mathrm{d}R.
\end{equation}
The key elements of this procedure are the PDF of the distance to the nearest base station $f_r(R)$ and the tail probability of the SINR conditioned on $r$, $\mathrm{P}\left[\gamma>y|r=R\right]$. The methodology to compute each of these elements while modelling the LOS and NLOS path loss components will be exposed next.
\vspace{-4mm}

\subsection{SPPPs of base stations in LOS and in NLOS with the user}

The set of the base stations locations originates an SPPP, which we denote by
$\Phi=\{x_n\}$.\footnote{Whenever there is no chance of confusion, we drop the
	subscript $n$ and use $x$ and instead of $x_n$ for convenience of notation.} As
a result of the propagation model we have adopted in our analysis (see
Section~\ref{subsect:ChannelModel}), the user can either be in LOS or NLOS with
any base station $x_n$ of $\Phi$.
Now, we perform the following mapping: we first define the set of LOS points,
namely $\Phi_{\mathrm{L}}$, and the set of NLOS points, $\Phi_{\mathrm{NL}}$.
Then, each point $x_n$ of $\Phi$ is mapped into $\Phi_{\mathrm{L}}$ if the base
station at location $x_n$ is in LOS with the user, while it is mapped to
$\Phi_{\mathrm{NL}}$ if the base station at location $x_n$ is in NLOS with the
user. Since the probability that $x_n$ is in LOS with the user is
$p_{\mathrm{L}}(\|x\|)$, it follows that each point $x_n$ of $\Phi$ is mapped
with probability $p_{\mathrm{L}}(\|x\|)$ into $\Phi_{\mathrm{L}}$ and
probability $p_{\mathrm{NL}}(\|x\|) = 1-p_{\mathrm{L}}(\|x\|)$ into
$\Phi_{\mathrm{NL}}$.
Given that this mapping is performed independently for each point in $\Phi$,
then from the "Thinning Theorem"~\cite[Theorem 2.36]{Haenggi2013} it follows
that the processes $\Phi_{\mathrm{L}}$ and $\Phi_{\mathrm{NL}}$ are SPPPs with
density $\lambda_{\mathrm{L}}(x)=\lambda p_{\mathrm{L}}(\|x\|)$
and $\lambda_{\mathrm{NL}}(x)=\lambda\left(1-p_{\mathrm{L}}(\|x\|)\right)$,
respectively. Note that, because of the dependence of $\lambda_{\mathrm{L}}(x)$
and $\lambda_{\mathrm{NL}}(x)$ on $x$, $\Phi_{\mathrm{L}}$ and
$\Phi_{\mathrm{NL}}$ are inhomogeneous SPPPs.
To make the formulation more tractable, we consider $\Phi_{\mathrm{L}}$ and
$\Phi_{\mathrm{NL}}$ to be independent processes; because the union of two SPPPs
processes is an SPPP of which the density is the sum of the densities of the
individual SPPPs \cite[Preposition 1.3.3]{Baccelli2009}, the union of $\Phi_{\mathrm{L}}$ and $\Phi_{\mathrm{NL}}$ is an SPPP with density $\lambda_{\mathrm{L}}(x)+\lambda_{\mathrm{NL}}(x)=\lambda$,
i.e., it is an SPPP with the same density as that of the original process
$\Phi$. Hence, the assumption of independence between $\Phi_{\mathrm{L}}$ and
$\Phi_{\mathrm{NL}}$ does not alter the nature of the process $\Phi$.

\vspace{-4mm}
\subsection{Mapping the NLOS SPPP into an equivalent LOS SPPP}
\label{subsect:Mapping}

Given that we have two inhomogeneous SPPP processes, it is not trivial to obtain the distribution of the minimum distance of the user to the serving base station, which will be necessary later on to compute the SINR CDF.
In fact, assuming the user be in LOS with the serving base station at a distance
$d_{\mathrm{1}}$, there might be an interfering BS at a distance
$d_\mathrm{2}<d_{\mathrm{1}}$ which is in NLOS with the user. This is possible
because the NLOS propagation is affected by a higher attenuation than the LOS
propagation.

Hence, to make our problem more tractable, we map the set of points of the
process $\Phi_{\mathrm{NL}}$, which corresponds to the NLOS base stations, into
an equivalent LOS process $\Phi_{\mathrm{eq}}$; each point
$x\in\Phi_{\mathrm{NL}}$ located at distance $d_{\mathrm{NL}}$ from the user is
mapped to a point $x_{\mathrm{eq}}$ located
at distance $d_{\mathrm{eq}}$ from the user, so that the BS located at
$x_{\mathrm{eq}}$ provides the same signal power to the user with path-loss
$K_{\mathrm{L}}d_{\mathrm{eq}}^{-\beta_{\mathrm{L}}}$ as if the base station
were located at $x$ with path-loss
$K_{\mathrm{NL}}d_{\mathrm{NL}}^{-\beta_{\mathrm{NL}}}$.

\begin{definition}[Mapping function
	$f_{\mathrm{eq}}$]\label{def:mapping_function}
	We define the mapping function
	$f_{\mathrm{eq}}:\Phi_{\mathrm{NL}}\rightarrow\Phi_{\mathrm{eq}}$ as:	
	\begin{equation}\label{eq:directMapping}
	f_{\mathrm{eq}}(x)=\frac{x}{\|x\|}d_{\mathrm{eq}}\left(\|x\|\right),
	\end{equation}
	\vspace{-3mm}
	\begin{equation}\label{eq:d_eq}
	d_{\mathrm{eq}}(d)=\left(\frac{K_{\mathrm{L}}}{K_{\mathrm{NL}}}\right)^{1/\beta_{\mathrm{L}}}d^{\beta_{\mathrm{NL}}/\beta_{\mathrm{L}}}.
	\end{equation}
\end{definition}
\begin{definition}[Inverse mapping function $g_{\mathrm{eq}}$]\label{def:inv_mapping_function}
	The inverse function
	$g_{\mathrm{eq}}=f_{\mathrm{eq}}^{-1}:\Phi_{\mathrm{eq}}\rightarrow\Phi_{\mathrm{NL}}$
	is defined as:
	\begin{equation}\label{eq:inverseMapping}
	\qquad	g_{\mathrm{eq}}(x)=\frac{x}{\|x\|}d_{\mathrm{eq}}^{-1}\left(\|x\|\right),
	\end{equation}
	\begin{equation}\label{eq:inverse_d_eq}
	d_{\mathrm{eq}}^{-1}(d)
	= \left(\frac{K_{\mathrm{NL}}}{K_{\mathrm{L}}}\right)^{1/\beta_{\mathrm{NL}}}
	d^{ \beta_{\mathrm{L}} / \beta_{\mathrm{NL}}}=
		K_{\mathrm{eq}} d^{\beta_{\mathrm{eq}}},
	\end{equation}
	where
	$K_{\mathrm{eq}}=\left(\frac{K_{\mathrm{NL}}}{K_{\mathrm{L}}}\right)^{1/\beta_{\mathrm{NL}}}$
	while $\beta_{\mathrm{eq}}=\beta_{\mathrm{L}}/\beta_{\mathrm{NL}}$.
\end{definition}
\vspace{-2mm}
It is important to notice that, from the "Mapping Theorem"~\cite[Theorem
2.34]{Haenggi2013}, $\Phi_{\mathrm{eq}}$ is still an SPPP.

\subsection{PDF of the distance from the user to the serving BS}\label{subsect:PDF_of_distance}

Using the mapping we introduced in Section \ref{subsect:Mapping}, we can
compute the PDF $f_{r}(R)$ of the minimum distance $r$ between the user and the
serving BS. We first compute the probability $\mathrm{P}\left[r>R\right]$; the
PDF can be ultimately obtained from the derivative of
$\mathrm{P}\left[r>R\right]$ as
$f_{r}(R)=\frac{\mathrm{d}}{\mathrm{d}R}(1-\mathrm{P}\left[r>R\right])
$. Let $B(0,l)$ be the ball of radius $l$ centred at the origin $(0,0)$. Moreover,
we use the notation $\Phi(\mathcal{A})$ to refer to number of points $x\in \Phi$
contained in $\mathcal{A}$~\cite{Haenggi2013}. Using the mapping we introduced
in Section \ref{subsect:Mapping} the probability $\mathrm{P}\left[r>R\right]$
can be found as:
\begin{equation*}
\mathrm{P}\left[r>R\right]=\mathrm{P}\left[
\Phi_{\mathrm{L}}\left(B(0,R)\right)=0\cap\Phi_{\mathrm{eq}}\left(B\left(0,R\right)\right)=0\right]
\end{equation*}
\begin{equation*}
\overset{(a)}{=}\mathrm{P}\left[\Phi_{\mathrm{L}}\left(B(0,R)\right)=0\cap\Phi_{\mathrm{NL}}\left(B\left(0,d_{\mathrm{eq}}^{-1}(R)\right)\right)=0\right]
\end{equation*}
\begin{equation}
\overset{(b)}{=}\mathrm{P}\left[\Phi_{\mathrm{L}}\left(B(0,R)\right)=0\right]\cdot\mathrm{P}\left[\Phi_{\mathrm{NL}}\left(B\left(0,d_{\mathrm{eq}}^{-1}(R)\right)\right)=0\right],
\end{equation}
where equality $(a)$ comes from the mapping defined in \eqref{eq:inverseMapping}
and in \eqref{eq:inverse_d_eq}, while equality $(b)$ comes from the independence
of the processes $\Phi_{\mathrm{L}}$ and $\Phi_{\mathrm{NL}}$. By applying the
probability function of inhomogeneous SPPP~\cite[Definition 2.10]{Haenggi2013},
we obtain the following,
\begin{equation}\label{eq:prob_r_greater_than_R}
\mathrm{P}\left[r>R\right]=\exp\bigg(-\int_{B(0,R)}\lambda_{\mathrm{L}}(x)\mathrm{d}x\bigg)
\exp\bigg(-\int_{B\left(0,d_{\mathrm{eq}}^{-1}(R)\right)}\lambda_{\mathrm{NL}}(x)\mathrm{d}x\bigg).
\end{equation}
From \eqref{eq:prob_r_greater_than_R}, we can obtain $f_{r}(R)$, first, by
integrating and, second, by computing its first derivative in $R$. The
formulation in~\eqref{eq:prob_r_greater_than_R} is general and thus can be
applied to several LOS probability functions $p_{\mathrm{L}}(d)$.
Below, we provide the expression of the PDF of the distance from the UE to the
serving BS for the LOS functions \eqref{eq:Our_p_L} and \eqref{eq:p_L_exp},
respectively.

\begin{result}\label{res:PDF_p_L_exp_square}
	If the LOS probability function is as in \eqref{eq:Our_p_L} and if we denote
	$d_{\mathrm{eq}}^{-1}(R)$ by $R_{\mathrm{eq}}$, the PDF of the distance to the
	serving BS is:
	\vspace{-1mm}
	\begin{align}\label{eq:PDF_exp_square}
	f_r(R)=-\left(e^{\pi\lambda L^{2}e^{-\frac{R^{2}}{L^{2}}}}\cdot e^{-\pi\lambda
		L^{2}e^{-\frac{R_{\mathrm{eq}}^{2}}{L^{2}}}}\cdot e^{-\pi\lambda
		R_{\mathrm{eq}}^{2}}\right)
	\end{align}
	\begin{align*}
	\left( -2\pi\lambda Re^{-\frac{R^{2}}{L^{2}}} \pi\lambda
	K_{\mathrm{eq}}^{2}2\beta_{\mathrm{eq}}R^{2\beta_{\mathrm{eq}}-1}e^{-\frac{-K_{e\mathrm{q}}^{2}R^{2\beta_{\mathrm{eq}}}}{L^{2}}}
	-\pi\lambda K_{\mathrm{eq}}^{2}2\beta_{\mathrm{eq}}R^{2\beta_{\mathrm{eq}}-1}
	\right).
	\end{align*}
\end{result}
\begin{result}\label{res:PDF_p_L_exp}
	If the LOS probability function is as in \eqref{eq:p_L_exp} and if we denote
	$d_{\mathrm{eq}}^{-1}(R)$ by $R_{\mathrm{eq}}$, the PDF of the distance to the
	serving BS is:
	\begin{align}\label{eq:PDF_exp}
	f_r(R)=-\left( e^{2\pi\lambda L^{2}e^{-\frac{R}{L}}}\cdot e^{2\pi\lambda LR
		e^{-\frac{R}{L}}}\cdot e^{-\pi\lambda R_{\mathrm{eq}}^{2}}\cdot e^{-2\pi\lambda
		L^{2}e^{-\frac{R_{\mathrm{eq}}}{L}}}\cdot e^{-2\pi\lambda
		LR_{\mathrm{eq}}e^{-\frac{R_{\mathrm{eq}}}{L}}} \right)
	\end{align}
\vspace{-7mm}
	\begin{align*}
	\bigg( -2\pi\lambda Le^{-\frac{R}{L}} -2\pi\lambda(L-R)e^{-\frac{R}{L}}
	-\pi\lambda K_{\mathrm{eq}}^{2}2\beta_{\mathrm{eq}}R^{2\beta_{\mathrm{eq}}-1}
	\end{align*}
\vspace{-7mm}
	\begin{align*}
	+ 2\pi\lambda LK_{\mathrm{eq}} \beta_{\mathrm{eq}} R^{\beta_{\mathrm{eq}}}
	e^{-\frac{K_{eq}R^{\beta_{\mathrm{eq}}}}{L}} + 2\pi\lambda
	LK_{\mathrm{eq}}\beta_{\mathrm{eq}}R^{\beta_{\mathrm{eq}}-1}(K_{\mathrm{eq}}R^{\beta_{\mathrm{eq}}}-L)e^{-\frac{K_{eq}R^{\beta_{\mathrm{eq}}}}{L}}
	\bigg).
	\end{align*}
\end{result}

\noindent We refer to the Appendix for the details of the $f_{r}(R)$ we have given in \eqref{eq:PDF_exp_square} and in \eqref{eq:PDF_exp}.

\vspace{-3mm}
\subsection{Spatial process of the interfering of the active base stations}
\label{subsec:partially_loaded_networks}

In this framework, we include the cases of partially loaded networks and of
frequency reuse and we treat them separately. First, we identify the set of BSs that are active, i.e., those having one or more users to serve. As we focus the analysis on the typical user, we can also identify the set of BSs that act as interferers for that user; an active BS (excluding the one serving the user) is seen as an interferer if that BS transmits over the same band used to serve that user.
In the following, we denote by
$\Phi_{\mathrm{A}}$ the set of active BSs, while we denote by $\Phi_{\mathrm{I}}$ the set of the interfering BSs.

Let us consider first the case of frequency reuse, in which all the base stations are active, but each of these only uses a portion of the spectrum, in order to reduce the interference. Since all the base stations are supposed to be active, the process $\Phi_{\mathrm{A}}$ is the same as $\Phi$. However, we assume each base station selects a channel in a random manner using, for instance, frequency ALOHA spectrum access \cite{Chandrasekhar2009}. With a frequency reuse factor of $N$, each base station uses $1$ out of $N$ channels, which is chosen independently from the other base stations. Hence, each BS interferes with a given user with probability $1/N$; this is equivalent to carrying out a thinning of the original process $\Phi$ with probability $1/N$; from the Thinning Theorem, we obtain that $\Phi_{\mathrm{I}}$ is a homogeneous process with density $\lambda_{\mathrm{I}}=\lambda/N$.

In regards to the partially loaded networks, we recall from Section
\ref{subsect:user_distribution} that a fraction of the base stations might be inactive and, as such, would not generate interference. Assuming all the BSs transmit over the same band, in partially loaded networks the active base stations are the only BSs that generate interference to the users---with the exception of the serving BS. Thus, we can write $\Phi_{\mathrm{I}} = \Phi_{\mathrm{A}} \setminus x_{0}$, where $x_{0}$ is the serving base station; moreover, from the Palm Theorem
\cite{Haenggi2013}, $\Phi_{\mathrm{I}}$ and $\Phi_{\mathrm{A}}$ are
characterized by the same density. To obtain the process of 	active BSs $\Phi_{\mathrm{A}}$  from the original the
process $\Phi$, we first assume that each UE deployed in the network connects to
the closest BS; \footnote{As we recall from Section \ref{subsect:Mapping}, with LOS/NLOS propagation the serving BS might not be the closest one to the user.
Nonetheless, this does not alter the validity of the explanation we are giving in this section.} finally, only the BSs which are assigned one or more users are active. Yet, this is equivalent to performing a thinning of the original process to obtain $\Phi_{\mathrm{A}}$. However, the fact that a base station is picked to be part of $\Phi_{\mathrm{A}}$ (i.e. there exists a user for which this BS is the closest one) depends on the positions of the neighbouring base stations, which implies that the base stations are not chosen independently of one another \cite{SeunghyunLee2012}. As the independence is one of the necessary conditions in order to have an SPPP, it follows that $\Phi_{\mathrm{A}}$ is not an SPPP.

Although $\Phi_{\mathrm{A}}$ cannot be formally regarded as an SPPP, it has been shown that the actual SPPP obtained through the thinning of $\Phi$ well
approximates $\Phi_{\mathrm{A}}$ \cite{SeunghyunLee2012,ChangLi2014}.
Specifically, the authors in \cite{SeunghyunLee2012} have shown that, (i) the probability $p_{\mathrm{A}}$ of a base station to be active (i.e., to have users to serve) can be well approximated once the density of users
$\lambda_{\mathrm{U}}$ and density of base stations $\lambda$ are known, and, (ii) the process $\Phi_{\mathrm{A}}$ of active base stations can be well
approximated by thinning the original process $\Phi_{\mathrm{A}}$ with
probability $p_{\mathrm{A}}$, which is given below \cite{SeunghyunLee2012}:
\begin{equation}\label{eq:p_active_BS}
p_{\mathrm{A}} =  1 - \left(1 +  \frac{\lambda _{\mathrm{U}}}{3.5 \lambda }
\right)^{-3.5}.
\end{equation}
From the Thinning Theorem, it follows that the resulting process obtained
through thinning as described above is an SPPP with density
$\lambda_{\mathrm{A}} = p_{\mathrm{A}}\lambda$; moreover, $\lambda_{\mathrm{I}}
= \lambda_{\mathrm{A}}$. In light of these findings, we approximate
$\Phi_{\mathrm{A}}$ with an SPPP of density $\lambda_{\mathrm{A}} =
p_{\mathrm{A}}\lambda$.

\begin{figure}[htbp]
	\centering
	\includegraphics[width=\figureSize\columnwidth]{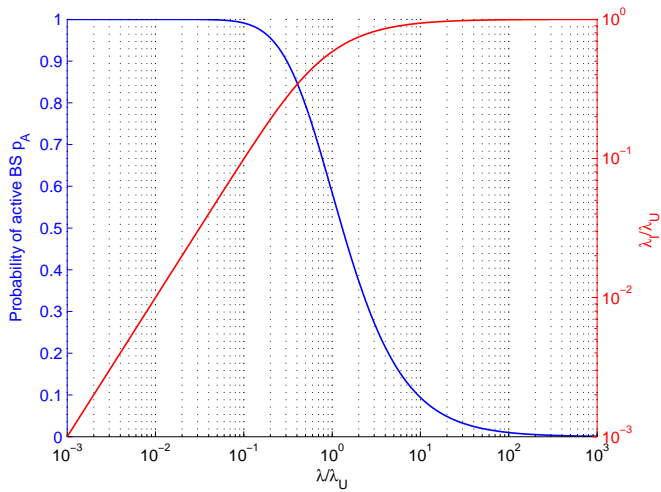}
	\caption{Probability of a BS being active and density of interfering BS vs BS density for partially loaded networks. The probability $p_{\mathrm{A}}$ drops as the ratio $\lambda / \lambda_{\mathrm{U}}$ is close to or greater than 1, i.e., as $\lambda$ approaches $\lambda_{\mathrm{U}}$. As a result of	 this, the density of active BSs as well as the density of interfering BSs converge to $\lambda_{\mathrm{U}}$ as $\lambda$ approaches or overcomes
$\lambda_{\mathrm{U}}$.}\label{fig:prob_A_vs_density}
\vspace{-7mm}
\end{figure}

Fig. \ref{fig:prob_A_vs_density} shows how the probability $p_{\mathrm{A}}$ and the intensity of  the interfering BSs $\lambda_{\mathrm{I}}$ vary as functions of the ratio $\lambda / \lambda_{\mathrm{U}}$.

\subsection{SINR inverse cumulative distribution function}\label{subsect:SINR_ICDF_Final}

The probability $\mathrm{P}\left[\gamma>y|r=R\right]$ can be computed as
in~\cite[Theorem 1]{Andrews2011}; we skip the details and provide the general formulation:
\begin{equation}\label{eq:P_SINR_geq_y_conditioned}
\mathrm{P}\left[\gamma>y|r=R\right]=\mathrm{P}\left[\frac{g
	K_{\mathrm{L}}R^{-\beta_{\mathrm{L}}}}{\sigma^{2}+I_{R}}>y\right]
=e^{-\mu y
	K_{\mathrm{L}}^{-1}R^{\beta_{\mathrm{L}}}\sigma^{2}}\mathcal{L}_{I_{R}}(\mu y
K_{\mathrm{L}}^{-1}R^{\beta_{\mathrm{L}}}),
\end{equation}
where $g$ is the Rayleigh fading, which we assume to be an exponential random variable $\sim exp(\mu)$; $\sigma^{2}$ is the variance of the additive white
Gaussian noise normalized with the respect to the transmit power,
$I_{R}$ is the interference conditioned on $R$, i.e., 	
\begin{equation}
I_{R} = \sum\limits _{\{i:\: x_i\in\Phi_{\mathrm{L}}\cap\Phi_{\mathrm{A}},\:
	\|x_i\|>R\}}g_i K_{\mathrm{L}}\|x_i\|^{-\beta_{\mathrm{L}}} + \sum\limits
_{\{j:\: f_{\mathrm{eq}}(x_j)\in \Phi_{\mathrm{NL}}\cap\Phi_{\mathrm{A}},\:
	\|f_{\mathrm{eq}}(x_j)\| > R\}}g_j K_{\mathrm{L}}\|x_j\|^{-\beta_{\mathrm{L}}}
\end{equation}
\begin{equation}
= \sum\limits _{\{i:\: x_i\in\Phi_{\mathrm{L}}\cap\Phi_{\mathrm{A}},\:
	\|x_i\|>R\}}g_i K_{\mathrm{L}}\|x_i\|^{-\beta_{\mathrm{L}}} + \sum\limits
_{\{j:\: x_j\in \Phi_{\mathrm{NL}}\cap\Phi_{\mathrm{A}},\: \|x_j\| >
	d_{\mathrm{eq}}^{-1}(R)\}}g_j K_{\mathrm{NL}}\|x_j\|^{-\beta_{\mathrm{NL}}}
\end{equation}
where $g_i$ and $g_j$ are independent and identically distributed $\sim
exp(\mu)$ fading coefficients. Please, note that the overall interference
accounts only for the active base stations. Compared to the formulation of $\mathcal{L}_{I_{R}}(s)$ proposed in \cite{Andrews2011}, in our case we have to deal with two non-homogeneous SPPP, namely $\Phi_{\mathrm{L}}$ and $\Phi_{\mathrm{NL}}$ instead of a single homogeneous SPPP.
The Laplace transform $\mathcal{L}_{I_{R}}(s)$ can be written
as follows:
\begin{align*}
\mathcal{L}_{I_{R}}(s)
=\mathrm{E}_{\Phi_{\mathrm{L}}\cap\Phi_{\mathrm{A}},\Phi_{\mathrm{NL}}\cap\Phi_{\mathrm{A}},g_i,g_j}\bigg[\exp\bigg(-s\sum\limits
_{\{i:\: x_i\in\Phi_{\mathrm{L}}\cap\Phi_{\mathrm{A}},\: |x_i|>|x_0|\}} g_i
K_{\mathrm{L}}\|x_i\|^{-\beta_{\mathrm{L}}}\bigg)\\
\exp\bigg(-s\sum\limits _{\{j:\: x_j\in
	\Phi_{\mathrm{NL}}\cap\Phi_{\mathrm{A}},\: \|x_j\| >
	d_{\mathrm{eq}}^{-1}(R)\}}g_j
K_{\mathrm{NL}}\|x_j\|^{-\beta_{\mathrm{NL}}}\bigg)\bigg].
\end{align*}
Given that $\Phi_{\mathrm{L}}$ and $\Phi_{\mathrm{NL}}$ are two independent
SPPP, we can separate the expectation to obtain:
\begin{align}\label{eq:laplace_notSolved}
\mathcal{L}_{I_{R}}(s)=\mathrm{E}_{\Phi_{\mathrm{L}}\cap\Phi_{\mathrm{A}},g_i}\bigg[\exp\bigg(-s\sum\limits_{\{i:\:
	x_i\in\Phi_{\mathrm{L}}\cap\Phi_{\mathrm{A}},\: \|x_i\|>R\}} g_i
K_{\mathrm{L}}\|x_i\|^{-\beta_{\mathrm{L}}}\bigg)\bigg] \\ \nonumber
\mathrm{E}_{\Phi_{\mathrm{NL}}\cap\Phi_{\mathrm{A}},g_j}\bigg[\exp\bigg(-s\sum\limits_{\{j:\:
	x_j\in \Phi_{\mathrm{NL}}\cap\Phi_{\mathrm{A}},\: \|x_j\| >
	d_{\mathrm{eq}}^{-1}(R)\}}g_j
K_{\mathrm{NL}}\|x_j\|^{-\beta_{\mathrm{NL}}}\bigg)\bigg].
\end{align}
By applying the Probability Generating Functional (PGFL) for SPPP (which holds also in case of inhomogeneous SPPP \cite{Haenggi2013}) to
\eqref{eq:laplace_notSolved}, we obtain the following result:
\begin{result} The Laplace transform $\mathcal{L}_{I_{R}}(s)$ for LOS/NLOS
	propagation with model given in \eqref{eq:propag_outdoor} is:
	\begin{equation*}
	\mathcal{L}_{I_{R}}(s)=\exp\Bigg(-2\pi\lambda_{\mathrm{I}}\int\limits
	_{R}^{+\infty}\left[\frac{sK_{\mathrm{L}}v^{-\beta_{\mathrm{L}}}}{sK_{\mathrm{L}}v^{-\beta_{\mathrm{L}}}+\mu}\right]p_{\mathrm{L}}(v)v\mathrm{d}v\Bigg)	
	\end{equation*}
	\begin{equation}\label{eq:laplace_solved}
	\exp\Bigg(-2\pi\lambda_{\mathrm{I}}\int\limits
	_{d_{\mathrm{eq}}^{-1}(R)}^{+\infty}\left[\frac{sK_{\mathrm{NL}}v^{-\beta_{\mathrm{L}}}}{sK_{\mathrm{NL}}v^{-\beta_{\mathrm{NL}}}+\mu}\right]p_{\mathrm{NL}}(v)v\mathrm{d}v\Bigg).
	\end{equation}

\end{result}
The Laplace transform  in \eqref{eq:laplace_solved} along with
\eqref{eq:prob_r_greater_than_R} and \eqref{eq:PDF_final_general} can be plugged in \eqref{eq:SINR_general_def} to obtain the SINR CCDF through numerical
integration.

\subsection{Average Spectral Efficiency and Area Spectral Efficiency}

Similarly to~\cite[Section IV]{Andrews2011} we compute the average spectral
efficiency and the ASE of the network. First, we define the ASE as:
\begin{equation}\label{eq:ASE}
\eta_{\mathrm{A}}=\frac{\lambda_{\mathrm{A}}\cdot A\cdot\mathrm{BW_{U}}\cdot
	\mathrm{E}[\mathrm{C}] }{A\cdot\mathrm{BW_{A}}}=\frac{\lambda_{\mathrm{A}}\cdot
	\mathrm{E}[\mathrm{C}] }{N},
\end{equation}
where $\mathrm{BW_{A}}$ is the available bandwidth, $\mathrm{BW_{U}}$ is the
used bandwidth, $\mathrm{E}[\mathrm{C}]$ is the average spectral efficiency, $A$
is the area, $N_{\mathrm{BS,A}}$ is the number of active base stations within
the area $A$, and $N$ is frequency reuse factor.
The average rate $\mathrm{E}[\mathrm{C}]$ can be computed as follows
\cite{Andrews2011}:
\begin{align}
\mathrm{E}[\mathrm{C}] = \mathrm{E}\left[\log_{2}(1+\gamma)\right]=\int
_{0}^{+\infty}\mathrm{P}\left[\log_{2}(1+\gamma)>u\right]\mathrm{d}u \\
\nonumber
=\int_{0}^{+\infty}\int
_{0}^{+\infty}\mathrm{P}\left[\log_{2}(1+\gamma)>u|r=R\right]f_{r}(R)\mathrm{d}R\mathrm{d}u.
\end{align}
\begin{equation}\label{eq:rate_final}
= \int _{0}^{+\infty}\int
_{0}^{+\infty}e^{-\mu(2^{u}-1)K_{\mathrm{L}}^{-1}R^{\beta_{\mathrm{L}}}\sigma^{2}}
\mathcal{L}_{I_{R}}\big(\mu(2^{u}-1)K_{\mathrm{L}}^{-1}R^{\beta_{\mathrm{L}}}\big)f_{r}(R)\mathrm{d}R\mathrm{d}u
\end{equation}
where $\mathcal{L}_{I_{R}}(s)$ is given in \eqref{eq:laplace_solved}. Similarly to the SINR CCDF, \eqref{eq:rate_final} can be evaluated numerically.


\vspace{-1mm}
\section{Energy efficiency with LOS/NLOS propagation} \label{sec:energy_Efficiency}

\subsection{Computing the transmit power per base station}\label{sub:Computing-the-transmit}

We evaluate the TX power in order to compute the overall power consumption
of the wireless network. Ideally, to ensure the network performance not be
limited by the transmit power, $P_{\mathrm{TX}}$ should be set in order to
guarantee the interference-limited regime, i.e.,  the transmit power should
be high enough so that the thermal noise power at the user receiver can be
neglected with respect to the interference power at the receiver.
In fact, when the network is in the interference-limited regime, the transmit power is high enough that any further increase of it would be pointless in
terms of enhancing the SINR, since the receive power increment would be
balanced by the exact same interference increment.

Practically, we refer to the outage probability $ \theta = \mathrm{P}\left[
\gamma \leq \gamma_{\mathrm{th}} \right]$ as a constraint to set the power
necessary to reach the interference limited regime. When the TX power is low, small increments of $P_{\mathrm{TX}}$ yields large improvements of the outage $\theta$; however, as $P_{\mathrm{TX}}$ increases, the corresponding outage
gain reduces, until $\theta$ eventually  converges to its optimal value $
\theta^*$, which would be reached in absence of thermal noise.  It is
reasonable to assume the network be in the interference-limited regime when
the following condition is met:
\vspace{-1mm}
\begin{equation}
| \theta^*-\theta | \leq \Delta_\theta,
\label{eq:Interf_limited_reg}
\end{equation}
where $\Delta_\theta$ is a tolerance measure setting the constraint in terms
of the maximum deviation of $\theta$ from the optimal value $\theta^*$. Eq. \eqref{eq:Interf_limited_reg} provides us with a metric to compute the
transmit power, but does not give us any indication on how to find $P_{\mathrm{TX}}$ as a function of the density $\lambda$. Unfortunately, we cannot derive a closed-form expression for the transmit power that satisfies \eqref{eq:Interf_limited_reg}, as we do not have any closed-form solution for the outage probability $ \theta = \mathrm{P}\left[ \gamma \leq \gamma_{\mathrm{th}} \right]$. We then take a different approach to calculate the minimum transmit power.

In Alg. \ref{algo:power} we proposed a simple iterative algorithm that finds
the minimum transmit power satisfying \eqref{eq:Interf_limited_reg} by using
numerical integration of \eqref{eq:SINR_general_def}. This algorithm computes the outage probability corresponding  to a given $P_{\mathrm{TX}}$; starting
from a low value of power, it gradually increases $P_{\mathrm{TX}}$ by a
given power step $\Delta_{P}$, until \eqref{eq:Interf_limited_reg} is
satisfied. To speed up this procedure, the step granularity is adjusted from
a coarse step $\mathrm{p}_1$ up to the finest step $\mathrm{p}_{N_{\mathbf{p
}}}$, which represents the precision of the power value returned by Alg. \ref{algo:power}.

\begin{algorithm}\caption{Steps to compute the transmit power.}\label{algo:power}
\begin{algorithmic}[0]
\small
\State INPUTS:
\vspace{-2mm}
\begin{enumerate}
	\item Vector of the power steps in dBm $\mathbf{p} = [\mathrm{p}_1,\ \cdots
\  \mathrm{p}_{N_{\mathbf{p}}}]$, $\mathrm{N}_{\mathbf{p}}$ is the length of
vector $\mathbf{p}$;
	\item Outage SINR threshold $\gamma_{\mathrm{th}}$;
	\item Outage tolerance $\Delta_{\theta}$;
\end{enumerate}

\State Initialize variables:
\vspace{-1mm}
\begin{enumerate}
	\item $P_{curr} = P_{N_0}$, where $P_{N_0}$ is the AWGN power in dBm over
the bandwidth $\mathrm{BW}_{\mathrm{U}}$
	\item $P_{\mathrm{fin}} = P_{curr}$
\end{enumerate}

\State Find optimal outage $\theta^* = \mathrm{P}\left[ \gamma \leq \gamma_{
\mathrm{th}} \right]$ by integrating \eqref{eq:SINR_general_def} with
parameter $\sigma^2=0$

\For{ $k = 1,\cdots,N_{\mathbf{p}}$}

\State  Find $\theta(P_{curr}) = \mathrm{P}\left[ \gamma \leq \gamma_{\mathrm{
th}} \right]$ by integrating \eqref{eq:SINR_general_def} with parameter $
\sigma^2=10^{-\frac{P_{curr}}{10}}$

\State Set granularity of the power step $\Delta_{P} = \mathbf{p}_k $

\While{$|\theta^*- \theta(P_{curr})|>\Delta_{\theta}$}

\State Increase the current power with step $\Delta_{P}$, i.e, $P_{curr} = P_{
curr} + \Delta_{P}$

\State Find $\theta(P_{curr}) = \mathrm{P}\left[ \gamma \leq \gamma_{\mathrm{
th}} \right]$ by integrating \eqref{eq:SINR_general_def} with parameter $
\sigma^2=10^{-\frac{P_{curr}}{10}}$

\State Update the final value of power, i.e., $P_{\mathrm{fin}} = P_{curr}$

\EndWhile

\State Remove the last power increment before increasing the granularity,
i.e.,  $P_{curr} = P_{curr} - \Delta_{P}$

\EndFor

\State OUTPUT: $P_{\mathrm{fin}}$ is the power in dBm s.t. \eqref{eq:Interf_limited_reg} is satisfied.
\end{algorithmic}
\end{algorithm}

\subsection{Energy efficiency}\label{sec:Energy_Efficiency}

In this subsection, we characterize the energy efficiency
of the network as a function of the base station density to identify
the trade-off between the area spectral efficiency and the power consumed
by network. We define the \textit{energy efficiency} as the ratio
between the overall throughput delivered by the network and the total
power consumed by the wireless network, i.e., we define the energy efficiency as follows:
\begin{equation}
\eta_{\mathrm{EE}}(\lambda)\triangleq\frac{T(\lambda)}{P_{\mathrm{TOT}}(
\lambda)},\label{eq:eff_definition}
\end{equation}
where $T(\lambda)$ is the network throughput, which can be written as $T(
\lambda)= A\cdot\mathrm{BW}\cdot \eta_{\mathrm{A}}(\lambda)$,
with \textbf{$\mathrm{BW}$} denoting the bandwidth and $\eta_{\mathrm{A}}(
\lambda)$ denoting the area spectral efficiency; $P_{\mathrm{TOT}}$ is the
overall power consumption of the network.

When we compute the power consumption of each BS, we need to take into
account that a fraction of the base stations may be inactive and model the
power consumption accordingly.  For active base station, we model the power
consumption $P_{\mathrm{BS,A}}$ of the base station assuming that $P_{\mathrm{
BS,A}}$ is the sum of two components, i.e., $P_{\mathrm{BS,A}}=P_{0}+P_{
\mathrm{RF}}$; the first, denoted by $P_{0}$,
takes into account the energy necessary for signal processing and
to power up the base station circuitry. This power $P_{0}$ is modelled
as a component being independent of the transmit power and of the
base station load \cite{Auer2011}. The second component, denoted
by $P_{\mathrm{RF}}$, takes into account the power fed into the power
amplifier which is then radiated for signal transmission. The power
$P_{\mathrm{RF}}$ is considered to be proportional to the power transmitted
by the base station; we can thus write  $P_{\mathrm{RF}}=K_{\mathrm{RF}}P_{
\mathrm{TX}}$, where $K_{\mathrm{RF}}$ takes into account the losses of the power
amplifier (i.e., we assume $K_{\mathrm{RF}}$ to be the inverse of the power
amplifier efficiency).

In the case of inactive base stations, we assume that the BS switches to a
stand-by state for energy saving purposes \cite{Ashraf2011}, in which it does not transmit (i.e., $P_{\mathrm{RF}}=0$) and reduces the circuitry power
consumption. Therefore, the power required to maintain the stand-by state can be modelled as $P_{\mathrm{BS,S}}=\rho P_0 $, where $\rho$ is power saving
factor that describes the relative power consumption of the circuitry with
respect to the active case; note that $0<\rho<1$.

Finally, the overall power consumption of the network due to both active and
inactive base station can be expressed as follows:
\begin{equation*}
P_{\mathrm{TOT}}= A \lambda_{\mathrm{A}} P_{\mathrm{BS,A}} + A(\lambda -
\lambda_{\mathrm{A}})P_{\mathrm{BS,S}}
\end{equation*}
\begin{equation}\label{eq:power_tot}
= A \lambda_{\mathrm{A}} P_0 + A \lambda_{\mathrm{A}} P_{\mathrm{TX}} K_{
\mathrm{RF}} + A (\lambda-\lambda_{\mathrm{A}})\rho P_0.
\end{equation}
The energy efficiency for the cases of fully loaded networks and partially
loaded networks is addressed in the following sub-sections.

\subsection{Energy efficiency for fully loaded networks} \label{subsec:energy_eff_fully_loaded}

In this section we study the energy efficiency $\eta_{\mathrm{EE}}(\lambda)$ trend  as a
function of $\lambda$; we focus on fully loaded networks, i.e., $p_{\mathrm{A}}=1$ and $\lambda_{\mathrm{A}}=\lambda$. Unfortunately, the analysis of the derivative of $\eta_{\mathrm{EE}}$
is not straightforward, as we have a closed-form solution neither for the
throughput $T(\lambda)$ nor for the transmit power $P_{\mathrm{TX}}(\lambda)$.
One feasible way to get around this burden is to approximate $T(\lambda)$ and
$P_{\mathrm{TX}}(\lambda)$ with functions in the form:
\begin{equation}
f(z)=az^b.\label{eq:linear_reg_func}
\end{equation}
The model in \eqref{eq:linear_reg_func} has two advantages: (i) it can be
easily derivated and, thus, is apt to investigate the existence of optima; (ii) it is well suited to fit the non-linear behaviour of ASE and TX power. In fact, we have shown in our previous work \cite{Galiotto2014} that both $T(\lambda)$ and $P_{\mathrm{TX}}(\lambda)$ can be approximated with a piece-wise function in the form \eqref{eq:linear_reg_func}, where the parameters $a$  and $b$ can be obtained, for instance, by linear regression in the logarithmic domain for a given range of values of $\lambda$.

Backed by the conclusions from previous work \cite{Galiotto2014}, we
approximate the throughput as $T(\lambda)=A T_{0}\lambda^{\alpha}$
and the transmit power as $P_{\mathrm{TX}}(\lambda)=P_{\mathrm{T}}\lambda^{
\delta}$, within a given interval of $\lambda$.
Under these assumptions, the energy efficiency becomes:
\begin{equation}
\eta_{\mathrm{EE}}(\lambda)=\frac{T_{0}\lambda^{\alpha}}{\lambda P_{0}+\lambda  K_{\mathrm{RF}}  P_{\mathrm{T}} \lambda^{\delta}}=\frac{T_{0}\lambda^{\alpha-1}}{P_{0}+ K_{\mathrm{RF}} P_{\mathrm{T}}\lambda^{\delta}}. \label{eq:eff_pathloss_full_load}
\end{equation}
The derivative of $\eta_{\mathrm{EE}}(\lambda)$ is given below:
\begin{equation}
\frac{\mathrm{d}\eta_{EE}(\lambda)}{\mathrm{d}\lambda}=\frac{T_{0}P_{0}(\alpha
-1)\lambda^{\alpha-2}+T_{0} K_{\mathrm{RF}} P_{\mathrm{T}}(\alpha-\delta-1)\lambda^{\alpha+
\delta-2}}{\left(P_{0}+K_{\mathrm{RF}} P_{\mathrm{T}}\lambda^{\delta}\right)^{2}}.
\end{equation}
Let us note that $T_{0}$, $P_{0}$, $K_{\mathrm{RF}}$ and $P_T$ are positive; moreover it is reasonable to assume that $\alpha>0$ (i.e., the area spectral efficiency is an increasing function of the density) and that $\delta<0$, i.e., the transmit power per BS is a decreasing function of the density. In the following paragraphs, we study the behaviour of the energy efficiency as function of the density $\lambda$ by analyzing the derivative $\eta^\prime_{\mathrm{EE}}(\lambda)$. We distinguish the following three cases.

\subsubsection{The energy efficiency is a monotonically increasing function}
\label{subsect:energy_eff_increasing}
This occurs if the ASE growth is linear or superlinear, i.e., if $\alpha\geq 1$. From this, if follows that also $\alpha\geq 1>1+\delta$ holds true; in
this case, $\eta^\prime_{\mathrm{EE}}(\lambda)$ is strictly positive, meaning that the energy efficiency increases as the density increases.

\subsubsection{The energy efficiency is a monotonically decreasing function}
\label{subsect:energy_eff_decreasing}

This occurs if the ASE growth is sublinear, i.e., if $\alpha<1$, and, in
addition, $\alpha<1+\delta$. Then, $\eta^\prime_{\mathrm{EE}}(\lambda)$ is
strictly negative and so the energy efficiency is a monotonically decreasing
function of the density $\lambda$.

\subsubsection{The energy efficiency exhibits an optimum point}	\label{subsect:energy_eff_maximum}

If ASE gain is sublinear (i.e. $\alpha<1$) but grows with a slope $\alpha$
sufficiently high, (i.e., $\alpha>1+\delta$), then we obtain that the
derivative $\eta\prime_{\mathrm{EE}}(\lambda)$
nulls for
\begin{equation}\label{eq:optimal_en_eff_point}
  \lambda_{0}=\left(\frac{P_{0}\left(1-\alpha\right)}{K_{\mathrm{RF}} P_{\mathrm{T}} \left(\alpha-\delta-1\right)}\right)^{1/\delta},
\end{equation}
is positive for $\lambda<\lambda_0$ and is negative for $\lambda>\lambda_0$. Therefore, $\lambda_0$ is a global maximum of the energy efficiency.

As a whole, the behavior of the spectral efficiency is due to how the growths of the ASE of the TX power relate among each other as the base station
density increases. If the ASE grows rapidly enough to counterbalance the
total power increase of the network given by the addition of base stations,
then the energy efficiency increases with the BS density; this means that
adding base station is profitable in terms of energy efficiency. Else, adding BSs turns  not to be profitable from energy efficiency point of view.

\subsection{Energy efficiency for partially loaded networks}\label{subsec:energy_eff_part_loaded}

For partially loaded networks, we only analyze the case where $\lambda >
\lambda_{\mathrm{U}}$, as the opposite case  of $\lambda <\lambda_{\mathrm{U
}}$ leads back to fully loaded networks. By using L'H\^{o}pital's rule, one
can show that \eqref{eq:p_active_BS} can be approximated by $p_{\mathrm{A}}
\cong \lambda_{\mathrm{U}}\lambda^{-1}$, for $\lambda$ is sufficiently
greater than $\lambda_{\mathrm{U}}$. By applying this approximation to \eqref{eq:power_tot}, we obtain:
\begin{equation}\label{eq:P_tot_approx_1}
P_{\mathrm{TOT}} = \lambda_{\mathrm{U}} P_0 (1-\rho) + \lambda \rho P_0 +
\lambda_{\mathrm{U}} K_{\mathrm{RF}} P_{\mathrm{T}} \lambda^{\delta}.
\end{equation}
It is known from \cite{Auer2011} that, as the BS density increases, the main
contribution to the total power consumption is due to the circuitry power $P_0$, while the transmit power becomes negligible for the overall power balance.
Therefore, to make the problem more tractable, we can further approximate the total power in \eqref{eq:P_tot_approx_1} as $P_{\mathrm{TOT}} \cong \lambda_{\mathrm{U}} P_0 (1-\rho) + \lambda \rho P_0$. From \eqref{eq:eff_definition},
by using the approximation $T(\lambda)=A T_{0}\lambda^{\alpha}$ for the
throughput and $P_{\mathrm{TOT}} \cong \lambda_{\mathrm{U}} P_0 (1-\rho) +
\lambda \rho P_0$ for the power, we obtain the following expression for the
energy efficiency:
\begin{equation}
\eta_{\mathrm{EE}}(\lambda)\cong \frac{T_{0}\lambda^{\alpha-1}}{ \lambda_{
\mathrm{U}} P_0 (1-\rho) + \lambda \rho P_0 }.\label{eq:eff_pathloss_part_load}
\end{equation}
To analyze the behaviour of the energy efficiency as a function of $\lambda$, we follow the same approach as in Section  \ref{subsec:energy_eff_fully_loaded}  and we compute the derivative of $\eta_{\mathrm{EE}}(\lambda)$, which is given below:
\begin{equation}
\frac{\mathrm{d}\eta_{EE}(\lambda)}{\mathrm{d}\lambda}=\frac{T_{0} \lambda^{
\alpha-1}\left( \lambda \rho (\alpha-1) +\alpha \lambda_{\mathrm{U}} (1-\rho )
  \right)}{ \left(\lambda_{\mathrm{U}} P_0 (1-\rho) + \lambda \rho P_0 \right)
^2}.
\end{equation}
As the ASE is known to be sub-linear for partially loaded networks \cite{Park2014,ChangLi2014}, we assume $0<\alpha <1$; moreover, the power saving factor $\rho$ satisfies $0<\rho <1$. Therefore, the derivative $\eta_{EE}^\prime$ nulls for:
\vspace{-3mm}
\begin{equation}\label{eq:optimal_lambda_partial_load}
\lambda^* = \frac{ \alpha \lambda_{\mathrm{U}} (1-\rho)  }{ \rho(1-\alpha )},
\end{equation}
it is positive for $\lambda < \lambda^*$ while it is negative for $\lambda >
\lambda^*$. Hence, $\lambda^*$ is a local maximum of the energy efficiency
for partially loaded networks and the energy efficiency decreases for
densities $\lambda > \lambda^*$. Note that, this result holds for
$\lambda$ sufficiently greater than $\lambda_{\mathrm{U}}$.

\vspace{-3mm}


\section{Results}\label{sect:results}

In this section we present and discuss the results we obtained by integrating numerically the expressions of outage probability, of the Spectral Efficiency (SE), and of the ASE. In Section \ref{subsect:rate_ASE_density}, \ref{sub:frequency_reuse} and \ref{sub:partially_loaded_results} we assume the network to be interference-limited (i.e., we set the thermal noise power to 0), while the noise is taken into account in Section \ref{sub:TXPowerPerBS_results} and \ref{sub:energy_efficiency_results}.

The parameters we used to obtain our results are specified in Table \ref{table:parameters}.
\vspace{-3mm}

\begin{table}[tbph]
\caption{Parameters for results section}
\label{table:parameters}
\centering%
\begin{tabular}{|p{4.5cm}|p{11cm}|}
\hline
\textbf{Parameter}  & \textbf{Value}\tabularnewline
\hline
Path-loss - Single slope & \multirow{1}{10cm}{$\mathrm{PL}_{\mathrm{SL}}(d_{\mathrm{km}}) = 140.7+36.7\log(d_{\mathrm{km}})$,~~$\beta=3.67$, ~~$K_{\mathrm{SL}}=10^{14.07}$  \cite{3GPP36814}}\tabularnewline
\hline
Path-loss - Combined LOS/NLOS & See \eqref{eq:propag_outdoor}; with $d$ in km, $K_{\mathrm{L}}=10^{10.38}$, $\beta_{\mathrm{L}}=2.09$,
$K_{\mathrm{NL}}=10^{14.54}$, $\beta_{\mathrm{NL}}=3.75$, $d_0 =0.156\mathrm{km}$, $d_1 = 0.03\mathrm{km}$ \, \cite{3GPP36814} \tabularnewline
\hline
Parameter $L$  & 82.5m, set so that (2) and (3) intersect at the point corresponding to probability 0.5 \tabularnewline
\hline
Bandwidth $\mathrm{BW}$ & 10 MHz centered at 2 GHz \tabularnewline
\hline
Noise  & Additive White Gaussian Noise (AWGN) with -174 dBm/Hz Power Spectral Density\tabularnewline
\hline
Noise Figure  & 9 dB\tabularnewline
\hline
Antenna at BS and UE  & Omni-directional with 0 dBi gain\tabularnewline
\hline
$K_{\mathrm{RF}}$ & $10$ \cite{Auer2011}\tabularnewline
\hline
$P_{0}$ & $10$W \cite{Auer2011}\tabularnewline
\hline
\end{tabular}
\vspace{-5mm}
\end{table}

\subsection{Spectral efficiency, outage probability and ASE}\label{subsect:rate_ASE_density}

In this subsection we assume the network to be fully loaded and with frequency reuse 1.  We compared the results for two LOS probability functions, namely \eqref{eq:Our_p_L} and \eqref{eq:p_L_exp}; we also compared the results for LOS/NLOS propagation with those obtained with a the single slope path-loss model.  We first analyze the outage probability (defined as $\theta = \mathrm{P}\left[ \gamma \leq \gamma_{\mathrm{th}} \right]$) results, which have been obtained by numerical integration of \eqref{eq:SINR_general_def}.

\vspace{-7mm}

\begin{figure}[htb]
\centerline{\subfloat[Outage vs base station density.]{\includegraphics[width
=0.52\columnwidth]{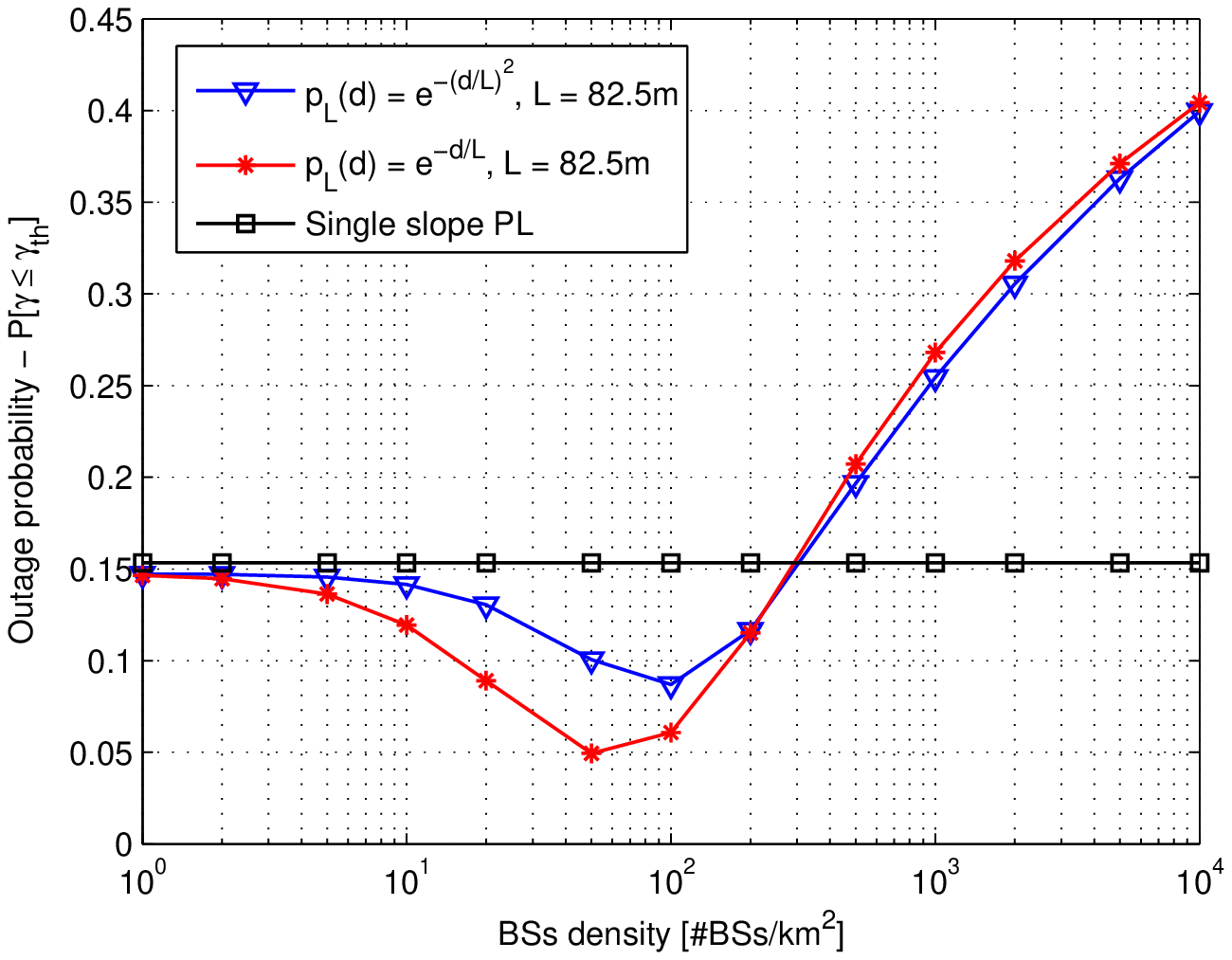}
\label{fig:outage_full_load}}
\hfil
\subfloat[ASE vs base station density.]{\includegraphics[width=0.52\columnwidth]{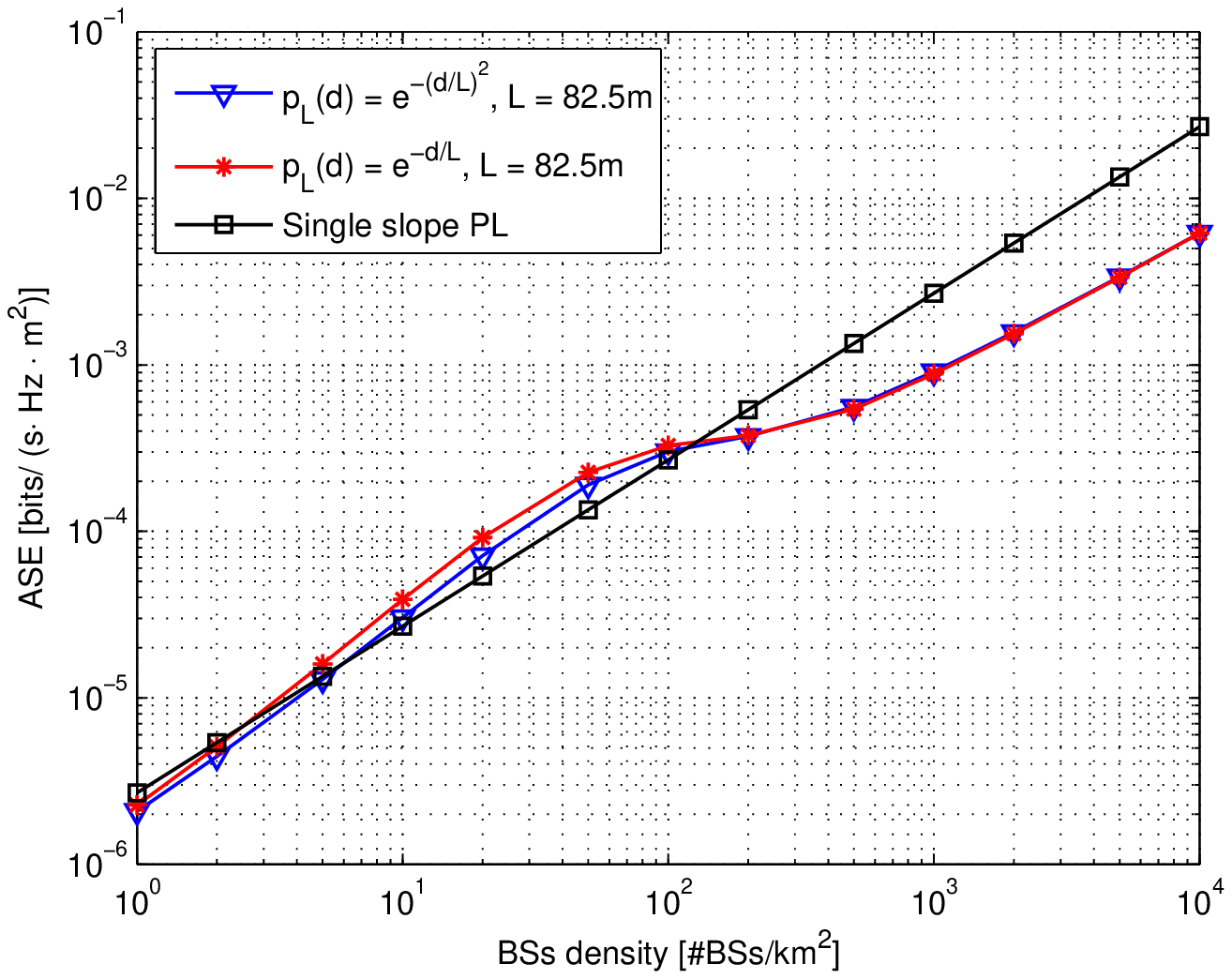}
\label{fig:ASE_full_load}}}
\caption{(a) Outage probability and (b) ASE vs base station density for different LOS probability function.}
\label{fig_sim}
\vspace{-6mm}
\end{figure}

We show the outage probability results in Fig. \ref{fig:outage_full_load}, where we can see the different effect of the LOS/NLOS propagation with respect to the single slope Path-Loss (PL). With single-slope PL, the outage is constant with the BS density. In contrast, with LOS/NLOS propagation, there is a minimum in the outage curves, which is achieved for density $\lambda$ = 50-100BSs/km$^2$, depending on the LOS probability function. Within this range of densities, the user is likely to be in LOS with the serving BS and in NLOS with most of the interfering BS, meaning that the interference power is lower than the received power.

At densities $\lambda$ greater than 300BSs/km$^2$, the outage starts growing drastically and, depending on the LOS likelihood, can reach 32-43\%. This is due to the fact that more and more interfering BSs are likely to enter the LOS region, causing an overall interference growth and thus a reduction of the SIR. At densities $\lambda$ smaller than 100BSs/km$^2$, the serving BS as well as the interfering BSs are likely to be in NLOS with the user. Because of this, both  the receive power and the overall interference increase at the same pace\footnote{If both serving BS and interfering BS are in NLOS with the user, the path-loss exponents of the serving BS-to-user channel and of the interfering BS-to-user channels are the same and, therefore, the power or the interference and of the received signal varies at the same slope as a function of the density.} and, as a consequence, the SIR remains constant, and so does the outage. Let us note that, the LOS probability function affects the outage curves at intermediate values of the BS density (e.g. 10-300 BSs/km$^2$). At low densities, all the BSs are likely to be in NLOS with the user, while at high  densities the serving BS and the strongest interferers are likely to be in LOS BSs are likely to be in LOS with the user.

The results of the ASE are shown in Fig. \ref{fig:ASE_full_load}. Compared to the single-slope PL, which shows a linear growth of the ASE with the density $\lambda$, with the LOS/NLOS propagation we observe a different behaviour of the ASE. In particular, we observe a lower steepness of the ASE curve at high BS densities, which is due to the effect of the interfering BSs entering the LOS region and, thus, increasing the total interference power.

To assess steepness of the ASE, we can use linear regression to interpolate the ASE curve with the model given in \eqref{eq:linear_reg_func}. In particular, we can approximate the ASE $\eta_{\mathrm{A}}(\lambda)$ with a piece-wise function of the kind  $\eta_{\mathrm{A}}(\lambda)=\eta_{\mathrm{A},0}\lambda^\alpha$, where $\eta_{\mathrm{A},0}$ and $\alpha$ are given for given intervals of $\lambda$. We specifically focus on $\alpha$, which gives the steepness of the ASE curve. With reference to the ASE curve (solid-blue curve in Fig. \ref{fig:ASE_full_load}) obtained with \eqref{eq:Our_p_L} as a LOS probability function, the the value of the parameter $\alpha$ turns to be 1.15 within the range of $\lambda$  1-50 BSs/km$^2$, 0.48 within the range 50-500 BSs/km$^2$ and 0.81 within the range 500-10000 BSs/km$^2$.

\subsection{Frequency reuse}\label{sub:frequency_reuse}

\begin{figure}[htb]
\centering
\includegraphics[width=0.52\columnwidth]{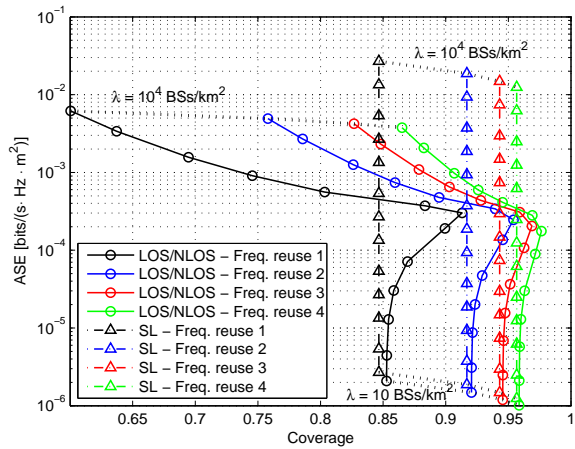}
\caption{ASE vs coverage trade-off for frequency reuse. The trade-off curves have been plotted for BS density equal to  10, 20, 50, 100, 200, 500, 1000, 2000, 5000, 10000 BSs/km$^2$, and compare the combined LOS/NLOS model with the single slope one.}\label{fig:ASE_coverage_tradeOff}
\vspace{-7mm}
\end{figure}

To have a comprehensive view of the frequency reuse as an interference mitigation scheme, we need to assess the trade-off between the ASE and the network coverage probability, defined as $1-\mathrm{P}\left[ \gamma \leq \gamma_{\mathrm{th}} \right]$. The results of this trade-off are shown in  Fig. \ref{fig:ASE_coverage_tradeOff}, where we plotted the network coverage against the ASE for different frequency reuse factors and base station densities.

Firstly, we focus on the LOS/NLOS propagation; we can notice from this plot that, if we fix the BS density, higher frequency reuse factors enhance the network coverage but, on the other hand, determine a drop of the ASE. This is in line with what one would expect from frequency reuse. Nonetheless, if we have no constraint in the choice of the BS density, the ASE vs coverage trade-off improves as the frequency reuse factor increases. In fact, the trade-off curve we obtain for a given reuse factor $K$ lies on the top-right hand side with respect to the curve for reuse factor $K-1$.  This means that, by increasing the reuse factor and the base station density at the same time, it is possible to achieve better performance than with a lower frequency reuse factors; note, though, that this is true when there is no constraint in terms of BS density.

By looking at the single slope PL curve in Fig. \ref{fig:ASE_coverage_tradeOff}, it appears that higher frequency reuse factors should still be preferred in order to improve the ASE vs coverage trade-off. However, unlike with the LOS/NLOS path loss, increasing the BS density enhances the ASE with no loss in terms of network coverage. Yet, modelling the signal propagation with the combined LOS/NLOS path loss yields different results than with the single-slope PL.
\vspace{-5mm}

\subsection{Partially loaded networks }\label{sub:partially_loaded_results}

\begin{figure}[!t]
\centerline{\subfloat[Outage vs base station density]{\includegraphics[width
=0.52\columnwidth]{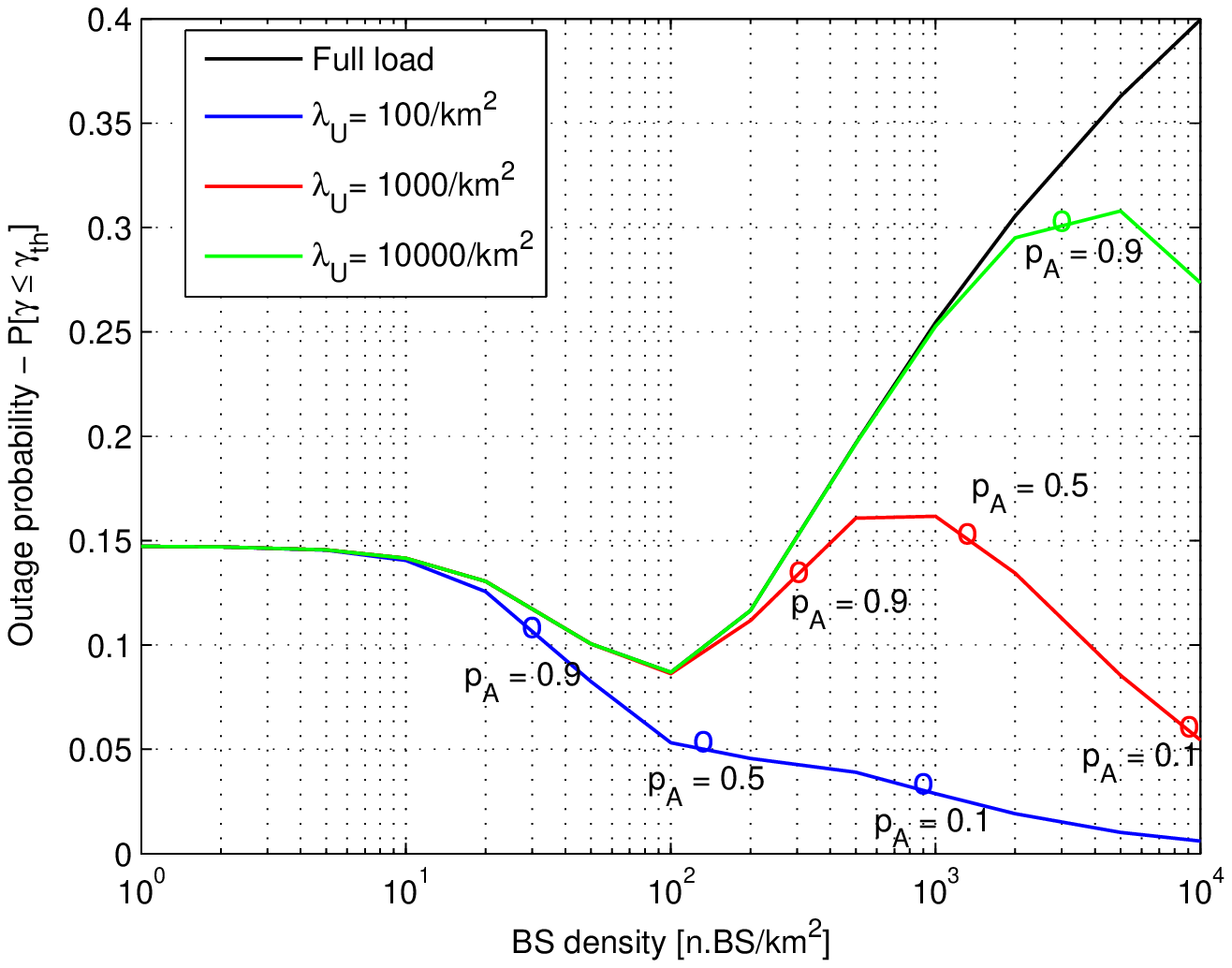}
\label{fig:outage_partial_load}}
\hfil
\subfloat[Spectral efficiency vs base station density.]{\includegraphics[width=0.52\columnwidth]{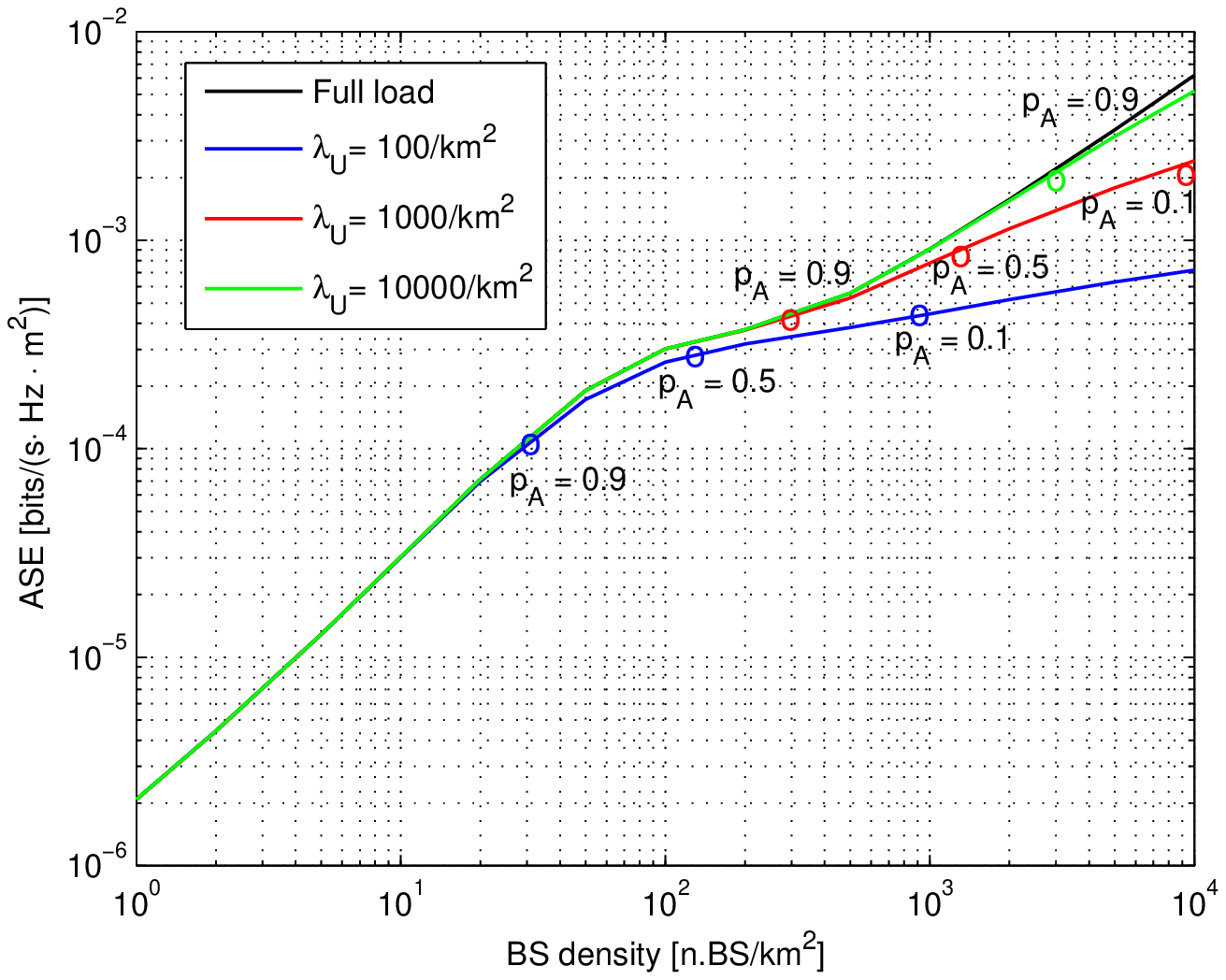}
\label{fig:ASE_partial_load}}}
\caption{The probability $p_{\mathrm{A}}$ given by \eqref{eq:p_active_BS} is reported as a dot on each curve. The outage probability has been obtained for $\gamma_{\mathrm{th}}=-8$dB.}
\label{fig_sim}
\vspace{-9mm}
\end{figure}

In this subsection we show the results of the cell densification for partially loaded networks with LOS/NLOS propagation. Differently from the case of fully loaded networks, we recall that a fraction of the BSs may be inactive and, thus, the density of interfering BSs $\lambda_{\mathrm{I}}$ does not necessary follow the trend of BS density $\lambda$ (see Section \ref{subsec:partially_loaded_networks} and Fig. \ref{fig:prob_A_vs_density}).
In Fig. \ref{fig:outage_partial_load} and \ref{fig:ASE_partial_load} we show the  outage probability and the ASE curves, respectively, as function of the base station density for difference user densities. To better understand the effect of the partial load on the network performance, we compare these curves with those for fully loaded networks. We reported the values of the probability $p_{\mathrm{A}}$ of a BS being active over the outage and ASE curves.

We observe that, as long as $p_{\mathrm{A}} \geq 0.9$, the deviation from the fully loaded network case is minimal. However, as soon as $\lambda$ approaches the value of user density $\lambda_{\mathrm{U}}$, the probability $p_{\mathrm{A}}$  drops and, as a consequence, the density of interfering $\lambda_{\mathrm{I}}$ BSs grows slowly with $\lambda$, up to the point where it saturates and converges to $\lambda_{\mathrm{U}}$ (see Fig. \ref{fig:prob_A_vs_density}).  At the same time, as $\lambda$ increases, the distance from UE to the serving BS tends to decrease, leading to an increment of the received power. Overall, the fact that $\lambda_{\mathrm{I}}$ saturates whereas the received power keeps growing as $\lambda$ increases has a positive impact on the SIR; as a result, the outage probability (see Fig. \ref{fig:outage_partial_load}) and the spectral efficiency improve once the density  $\lambda$ approaches or overcomes $\lambda_{\mathrm{U}}$.

In regards to the ASE trend, we show the results in Fig. \ref{fig:ASE_partial_load}. According to \eqref{eq:ASE}, the ASE trend is the combined outcome of the increase of the spectral efficiency and of the density of the active base stations. As the density of base stations increases and approaches the user density $\lambda_{\mathrm{U}}$, the density of active base stations will converge to $\lambda_{\mathrm{U}}$ (see Fig. \ref{fig:outage_partial_load}); given that the density of active BSs remains constant, the only contribution to the ASE increase will be given by the spectral efficiency improvement. As a matter of fact, we can see that, with respect to the case of fully loaded networks, the ASE curves show a lower gain when the density $\lambda$ approaches $\lambda_{\mathrm{U}}$.

To assess steepness of the ASE, we applied linear regression to the ASE curves in order to obtain the value of the parameter $\alpha$  corresponding to different intervals of $\lambda$;  we specifically consider the approximation for the curve corresponding to $\lambda_{\mathrm{U}}=1000$UEs/km$^2$ (red curve in Fig. \ref{fig:ASE_partial_load}). These values are $\alpha = 1.15 $ within the density range 1-50 BSs/km$^2$, $\alpha = 0.43 $ within the density range 50-500 BSs/km$^2$ and $\alpha = 0.46 $ within the density range 500-10000 BSs/km$^2$.

\subsection{Transmit power per base station }\label{sub:TXPowerPerBS_results}

\begin{figure}
\centering
\includegraphics[width=0.52\columnwidth]{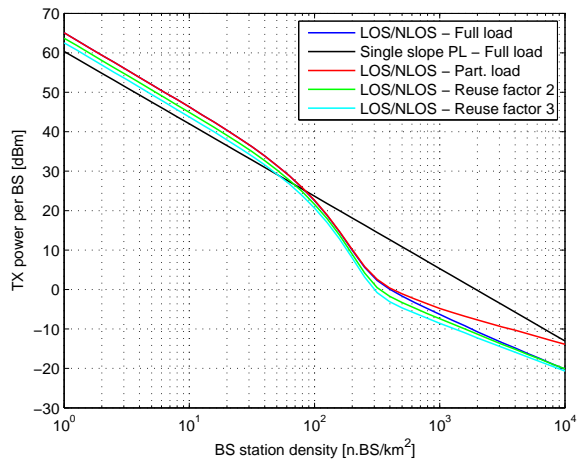}
\caption{Transmit power per BS. The power has been obtained with an SINR threshold $\gamma_{\mathrm{th}}=-8$dB and for tolerance $\Delta_{\theta}= 0.1\%$. The plot compares the TX power per BS for single slope-slope and  LOS/NLOS path-loss for fully loaded networks. It also provides the curve for frequency reuse factor 2 and 3 and for partially loaded network with $\lambda_{\mathrm{U}}=1000$UEs/km$^2$.}\label{fig:transmit_power}
\vspace{-9mm}
\end{figure}

In Fig. \ref{fig:transmit_power} we show the simulation results of
the transmit power per base station $P_{\mathrm{TX}}(\lambda)$, which has been computed by using Algorithm \ref{algo:power} as explained in Section
\ref{sub:Computing-the-transmit}. In this figure we compare the results
we obtained using the \textit{single slope} and the \textit{combined LOS/NLOS}
path loss models.

As we can see from this plot, the behaviour of the transmit power as a function of the BS density $\lambda$ is different in the two cases of single slope and combined LOS/NLOS propagation. With reference to Fig. \ref{fig:transmit_power}, with single slope path loss, the power decreases linearly (in logarithmic scale) with the density; in the case of combined LOS/NLOS propagation, the transmit power exhibits different slopes as the as the base station density increases. We used linear regression to assess the slopes  of the TX power curves (indicated by $\delta$, as explained in Section \ref{subsec:energy_eff_fully_loaded}) within different density intervals. With reference to the curve corresponding to fully loaded networks with LOS/NLOS propagation (solid-blue curve in Fig. \ref{fig:transmit_power}), the values of ($P_\mathrm{T}$, $\delta$) are $(9.3\cdot 10^{-9}, -1.9)$ within the $\lambda$ range 1-60 BSs/km$^2$, $(4.4\cdot 10^{-17}, -3.9)$ within the $\lambda$ range 60-300 BSs/km$^2$ and $(1.15\cdot 10^{-9}, -1.44)$ within the range 300-10000BSs/km$^2$.

The fact that the transmit power per base station decays more or less steeply with the density $\lambda$ depends on how quickly the interference power increases or decreases with $\lambda$. As we explained in Section \ref{sub:Computing-the-transmit}, the transmit power per base station $P_{\mathrm{TX}}(\lambda)$ has to be set so that the network is interference limited. Thus, if the channel attenuation between the interferer and the user decreases quickly as the density increases, a lower transmit power will be enough to guarantee that the interference power is greater than the noise power.
In other words, if the interferer-to-user channel attenuation tends to decrease quickly as the density increases, so does the transmit power and vice-versa.
For instance, for $\lambda \in[60,300]$BSs/km$^2$, the probability of having interferers in LOS with the user rises and, as a consequence, we have a lower attenuation of the channel between the interfering base station and the user.
Hence, the $P_{\mathrm{TX}}(\lambda)$ which  guarantees the interference-limited regime will also decrease steeply  with $\delta=-3.9$ as $\lambda$ increases. On the contrary, for $\lambda>300$ BSs/km$^{2}$, most of the interferers will have already entered the LOS zone, meaning that the interferer-to-user channel attenuation drops less rapidly than for $\lambda<300$ BSs/km$^{2}$; for this reason, also $P_{\mathrm{TX}}(\lambda)$ will decrease less rapidly with $\delta = -1.44$.

Let us note that, with increasing reuse factors $N$, the TX power decreases, as indeed a smaller bandwidth is used and, thus, the noise power is lower.

\subsection{Energy efficiency }\label{sub:energy_efficiency_results}

\begin{figure}[!t]
\centerline{\subfloat[Energy efficiency for fully loaded networks.]{\includegraphics[width
=0.55\columnwidth]{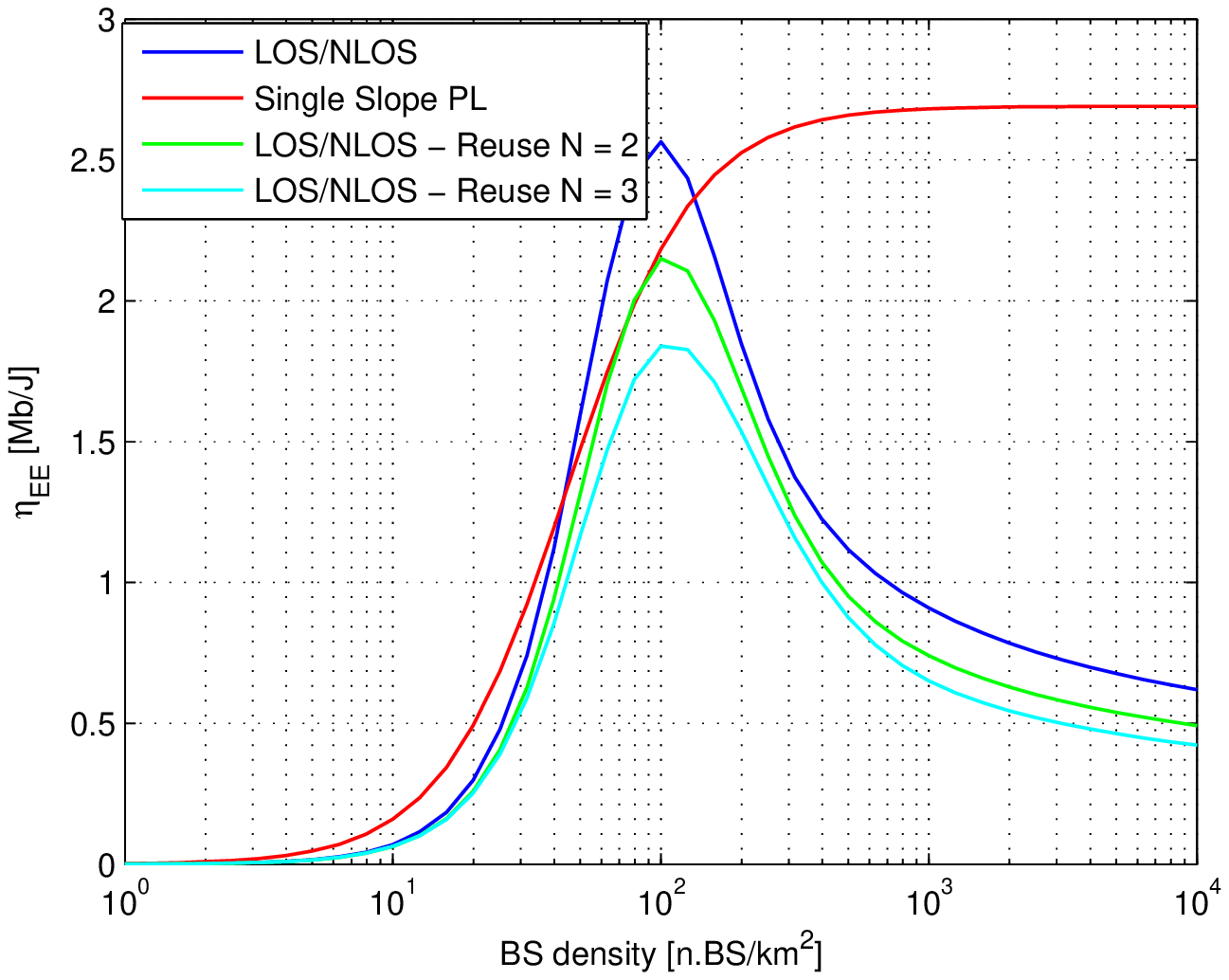}
\label{fig:energy_efficiency_v1}}
\hfil
\subfloat[Energy efficiency for partially loaded networks.]{\includegraphics[width=0.55\columnwidth]{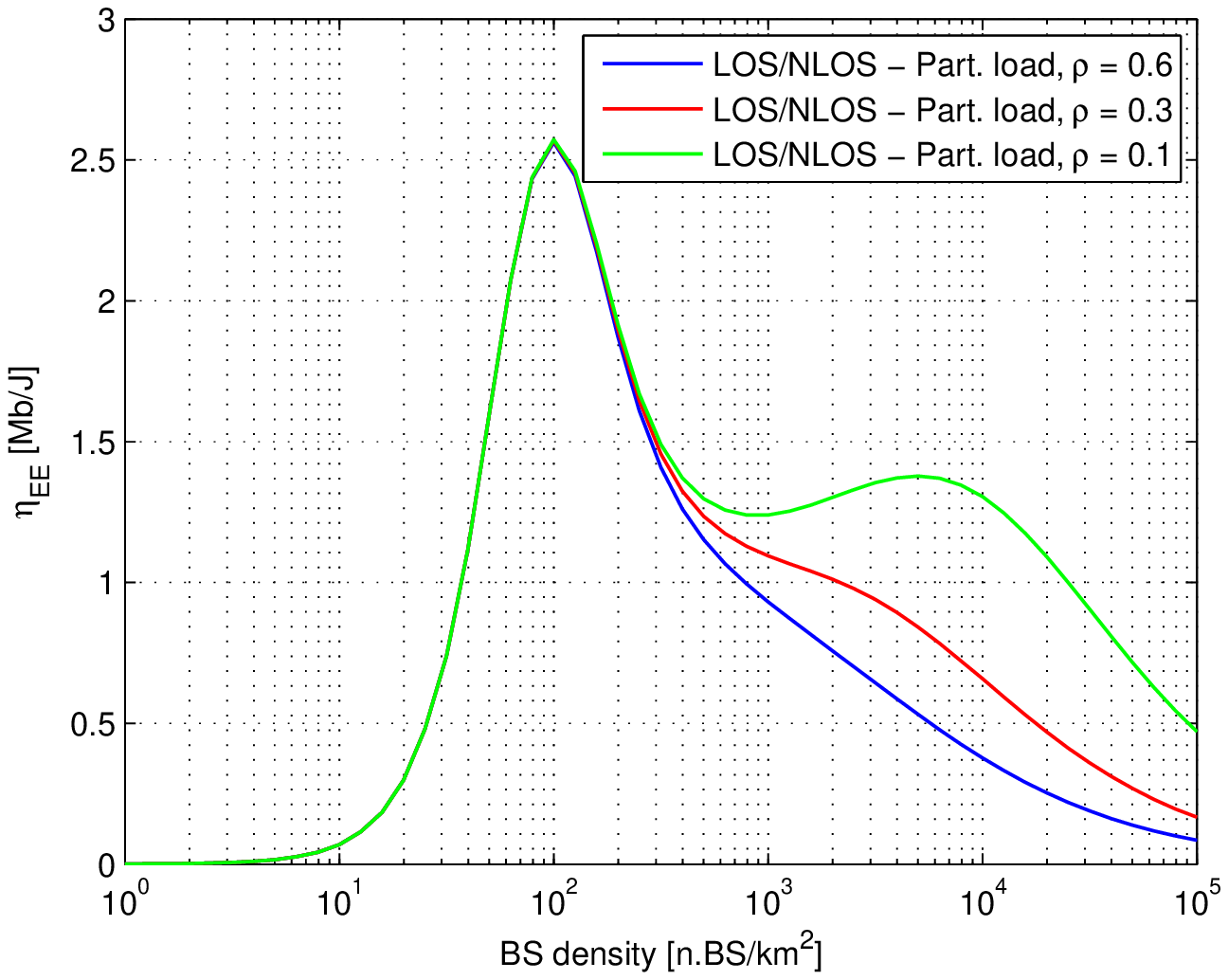}
\label{fig:energy_efficiency_part_load}}}
\caption{(a) Energy efficiency vs BS density for fully loaded networks. The plot compares the energy efficiency for LOS/NLOS with single slope path-loss. The energy efficiency is given also for frequency reuse factors 2 and 3. (b) Energy efficiency vs BS density for partially loaded networks. Curves are given for $\lambda_{\mathrm{U}}=1000$UEs/km$^2$ and for three values of $\rho$.}
\label{fig_sim}
\vspace{-9mm}
\end{figure}

One of the most surprising outcomes of our study on LOS/NLOS propagation for ultra-dense networks is the effect of cell-densification on the energy efficiency of the fully loaded network, of which we show the results in Fig. \ref{fig:energy_efficiency_v1}.
The difference between the energy efficiency with single-slope and with LOS/NLOS path-loss is noticeable. In the case of single-slope PL, due to the linear growth of the ASE, $\eta_{\mathrm{EE}}(\lambda)$ is a monotonically increasing function of the density $\lambda$ (see  Section \ref{subsect:energy_eff_increasing}). In the case of LOS/NLOS propagation, from Fig. \ref{fig:energy_efficiency_v1} we observe that the energy efficiency exhibits a maximum, which is achieved for a given density $\lambda_0$.

To explain this, we consider the case frequency reuse $N=1$ (solid-blue curve in Fig. \ref{fig:energy_efficiency_v1}); from \eqref{eq:optimal_en_eff_point} and with the values of the parameters $P_0$ (given in Table  \ref{table:parameters}), $P_\mathrm{T}$ and $\delta$ (given in Section \ref{sub:TXPowerPerBS_results}), and $\alpha$ (given in Section \ref{subsect:rate_ASE_density}), the optimal point $\lambda_0$ is approximately 100BSs/km$^2$. Beyond this point, the ASE gain is too low to compensate power consumption increase in the network, leading to a drop in terms of energy efficiency. From Fig. \ref{fig:energy_efficiency_v1}, we can note that frequency reuse reduces the energy efficiency compared to $N=1$. As a result of the lower ASE achieved at higher frequency reuse factors $N$, the energy efficiency drops as $N$ increases.

In Fig. \ref{fig:energy_efficiency_part_load} we show the energy efficiency for partially loaded networks, for a user density $\lambda_{\mathrm{U}}$ of 1000 UEs/km$^2$. As we are dealing with partially loaded networks, we are interested in the BS densities $\lambda>\lambda_{\mathrm{U}}$, where energy efficiency strongly depends on the power saving factor $\rho$ of the BSs in stand-by state. This is because the parameter  $\rho$ determines the energy saving of the inactive BSs, which become more numerous as the density $\lambda$ increases.  Depending on the value  of $\rho$, according to \eqref{eq:optimal_lambda_partial_load} a local maximum may even occur at $\lambda^* = \frac{ \alpha \lambda_{\mathrm{U}} (1-\rho)  }{ \rho(1-\alpha )}$.

With $\rho=0.1$ and with the values of $\alpha$ given in Section \ref{sub:partially_loaded_results}, the local maximum turns to be $\lambda^*\cong 7300$BSs/km$^2$. For higher values of $\rho$, $\lambda^*$ is smaller than or too close to $\lambda_{\mathrm{U}}$ to be considered as a reliable estimate of a maximum; we  recall from Section \ref{subsec:energy_eff_part_loaded} that this estimate can be reckoned as reliable only if $\lambda^*$ is sufficiently greater than $\lambda_{\mathrm{U}}$. In fact, we observe from Fig. \ref{fig:energy_efficiency_part_load} that there is no local maximum beyond $\lambda_{\mathrm{U}}$ for $\rho=0.3$ or $0.6$.


\vspace{-3mm}

\section{Conclusions}\label{sect:conclusions}

In this paper, we have proposed a stochastic geometry-based framework to model the outage probability, the Area Spectral Efficiency (ASE) of fully loaded and partially loaded Ultra-Dense Networks (UDNs), where the signal propagation accounts for LOS and NLOS components. We also studied the energy efficiency of UDNs resulting from this propagation model.

As the main findings of our work, we have shown that, with LOS/NLOS propagation, massive cell densification determines a deterioration of the network coverage at high cell densities, if the network is fully loaded. Moreover, the ASE grows less steeply than a linear function at high cell densities, which implies that a larger number of base stations would be required to achieve a given throughput target with respect to the case of single slope path-loss. In regards to the energy efficiency, cell densification turns out to be inefficient for the network from an energetic point of view.
In partially loaded networks, when the base station density exceeds that of the users, cell densification results in a coverage improvement. Overall, based on our findings, we can conclude that UDNs are likely face coverage issues in highly crowded environments with many users, which represent the worst case scenario for ultra-dense networks.

\vspace{-3mm}

\appendices{
\numberwithin{equation}{section}

\section{PDF of the distance to the serving BS}

Once the LOS probability function is known, from \eqref{eq:prob_r_greater_than_R} we obtain the PDF of the distance to the closest BS as follows:
\begin{equation}\label{eq:Prob_r_gt_R_integral}
\mathrm{P}\left[r>R\right]=\exp\bigg(-\lambda\int_{B(0,R)}p_{\mathrm{L}}(\|x\|)\mathrm{d}x\bigg)
\exp\bigg(-\lambda\int_{B\left(0,d_{\mathrm{eq}}^{-1}(R)\right)}\left(1-p_{\mathrm{L}}(\|x\|)\right)\mathrm{d}x\bigg).
\end{equation}
Assuming the integrals in \eqref{eq:Prob_r_gt_R_integral} can be solved in a closed-form,  with some symbolic manipulation, \eqref{eq:Prob_r_gt_R_integral} solves in its general form as follows:
\begin{equation}\label{eq:Prob_r_gt_R_integral_general}
\mathrm{P}\left[r>R\right]=\prod_{m=1}^{M}\exp(f_m(R)).
\end{equation}
By taking the derivative of \eqref{eq:Prob_r_gt_R_integral_general}, we obtain:
\vspace{-2mm}
\begin{equation*}
\frac{\mathrm{d}}{\mathrm{d}R}\left[\mathrm{P}\left[r>R\right]\right]=\frac{\mathrm{d}}{\mathrm{d}R}\left[\prod_{m=1}^{M}\exp(f_m(R))\right]=\sum_{m=1}^{M}\frac{\mathrm{d}}{\mathrm{d}R}\left[\exp(f_m(R))\right]\prod_{n=1,n\neq m}^{M}\exp(f_n(R))=
\end{equation*}
\vspace{-2mm}
\begin{equation*}
\sum_{m=1}^{M}\frac{\mathrm{d}}{\mathrm{d}R}\left[f_m(R)\right]\exp(f_m(R))\prod_{n=1,n\neq m}^{M}\exp(f_n(R))=\sum_{m=1}^{M}f_m^{\prime}(R)\prod_{n=1}^{M}\exp(f_n(R))=
\end{equation*}
\vspace{-2mm}
\begin{equation}\label{eq:derivative_general}
\sum_{m=1}^{M}f_m^{\prime}(R)\left(\prod_{n=1}^{M}\exp(f_n(R))\right)=\mathrm{P}\left[r>R \right] \sum_{m=1}^{M}f_m^{\prime}(R).
\end{equation}
The PDF of the distance to the serving BS can finally be obtained as
\begin{equation}\label{eq:PDF_final_general}
f_{r}(R) =- \frac{\mathrm{d}}{\mathrm{d}R}\left[\mathrm{P}\left[r>R\right]\right] = - \mathrm{P}\left[r>R \right] \sum_{m=1}^{M}f_m^{\prime}(R).
\end{equation}
If we assume the LOS probability to be given by~\eqref{eq:Our_p_L},
we can further develop \eqref{eq:Prob_r_gt_R_integral} by solving the integrals in \eqref{eq:Prob_r_gt_R_integral} and, with further symbolic manipulation, we obtain:
\begin{equation}\label{eq:CDF_exp_square}
\mathrm{P}\left[r>R\right]=e^{\pi\lambda L^{2}e^{-\frac{R^{2}}{L^{2}}}}\cdot e^{-\pi\lambda L^{2}e^{-\frac{R_{\mathrm{eq}}^{2}}{L^{2}}}}\cdot e^{-\pi\lambda R_{\mathrm{eq}}^{2}},
\end{equation}

\vspace{-2mm}
\noindent where $R_{\mathrm{eq}} = d_{\mathrm{eq}}^{-1}(R)$. Let us define the functions $f_{1}(R)$, $f_{2}(R)$, $f_{3}(R)$ and their first derivatives  $f_{1}^{\prime}(R)$,
$f_{2}^{\prime}(R)$, and $f_{3}^{\prime}(R)$, respectively, as follows:
$$
f_{1}(R)=\pi\lambda L^{2}e^{-\frac{R^{2}}{L^{2}}},\quad f_{2}(R)=-\pi\lambda L^{2}e^{-\frac{R_{\mathrm{eq}}^{2}}{L^{2}}}, \quad f_{3}(R)=-\pi\lambda R_{\mathrm{eq}}^{2},
\quad
f_{1}^{\prime}(R)=-2\pi\lambda Re^{-\frac{R^{2}}{L^{2}}},
$$
$$
f_{2}^{\prime}(R)=\pi\lambda K_{\mathrm{eq}}^{2}2\beta_{\mathrm{eq}}R^{2\beta_{\mathrm{eq}}-1}e^{-\frac{-K_{e\mathrm{q}}^{2}R^{2\beta_{\mathrm{eq}}}}{L^{2}}}, \qquad
f_{3}^{\prime}(R)=-\pi\lambda K_{\mathrm{eq}}^{2}2\beta_{\mathrm{eq}}R^{2\beta_{\mathrm{eq}}-1}.
$$
By plugging \eqref{eq:CDF_exp_square} and  $f_{1}^{\prime}(R)$,
$f_{2}^{\prime}(R)$, and $f_{3}^{\prime}(R)$ in \eqref{eq:PDF_final_general}, we obtain the PDF of the distance to the serving BS.
%
%


When the LOS probability function is given by \eqref{eq:p_L_exp}, we obtain the PDF of distance to the closest BS station as follows. First, by solving the integrals in \eqref{eq:Prob_r_gt_R_integral} and by some additional algebraic operations, we obtain $\mathrm{P}\left[r>R\right]$ as follows:
\begin{equation}\label{eq:CDF_exp}
\mathrm{P}\left[r>R\right]=e^{2\pi\lambda L^{2}e^{-\frac{R}{L}}}\cdot e^{2\pi\lambda LRe^{-\frac{R}{L}}}\cdot e^{-\pi\lambda R_{\mathrm{eq}}^{2}}\cdot e^{-2\pi\lambda L^{2}e^{-\frac{R_{\mathrm{eq}}}{L}}}\cdot e^{-2\pi\lambda LR_{\mathrm{eq}}e^{-\frac{R_{\mathrm{eq}}}{L}}}.
\end{equation}
Then, we define the functions $f_{1}(R),\,f_{2}(R)\cdots,\,f_{5}(R)$ and we compute their respective derivatives $f_{1}^{\prime}(R),\,f_{2}^{\prime}(R)$ $\cdots,\,f_{5}^{\prime}(R)$ as follows:
$$
f_1(R) = 2\pi\lambda L^{2}e^{-\frac{R}{L}},\quad f_{1}^{\prime}(R)=-2\pi\lambda Le^{-\frac{R}{L}},\quad
f_2(R) = 2\pi\lambda LRe^{-\frac{R}{L}},\quad f_{2}^{\prime}(R)=-2\pi\lambda(L-R)e^{-\frac{R}{L}},
$$
$$
f_3(R) = -\pi\lambda R_{\mathrm{eq}}^{2},\quad f_{3}^{\prime}(R)=-\pi\lambda K_{\mathrm{eq}}^{2}2\beta_{\mathrm{eq}}R^{2\beta_{\mathrm{eq}}-1},	\quad
f_4(R) = -2\pi\lambda L^{2}e^{-\frac{R_{\mathrm{eq}}}{L}},
$$
$$
f_{4}^{\prime}(R)=2\pi\lambda LK_{\mathrm{eq}}\beta_{\mathrm{eq}}R^{\beta_{\mathrm{eq}}}e^{-\frac{K_{eq}R^{ \beta_{\mathrm{eq}}}}{L}},\quad
f_5(R) = -2\pi\lambda LR_{\mathrm{eq}}e^{-\frac{R_{\mathrm{eq}}}{L}},$$
$$
f_{5}^{\prime}(R)=2\pi\lambda LK_{\mathrm{eq}}\beta_{\mathrm{eq}}R^{\beta_{\mathrm{eq}}-1}(K_{\mathrm{eq}} R^{\beta_{\mathrm{eq}}}-L)e^{-\frac{K_{eq}R^{\beta_{\mathrm{eq}}}}{L}},
$$
Finally, the PDF can be obtained by plugging $f_{1}^{\prime}(R),\,f_{2}^{\prime}(R)$ $\cdots,\,f_{5}^{\prime}(R)$ and \eqref{eq:CDF_exp} in \eqref{eq:PDF_final_general}.

}


\end{document}